\documentclass[11pt]{article}
\usepackage[margin=1in]{geometry}
\usepackage[utf8]{inputenc}			
\usepackage{bbm}
\usepackage[labelfont=sc]{caption}
\usepackage[singlespacing]{setspace}
\usepackage[font=small,labelfont=normalfont,subrefformat=parens]{subcaption}
\usepackage[usenames,svgnames]{xcolor}
\usepackage[bookmarks=true,colorlinks=true,allcolors=MidnightBlue]{hyperref}

\usepackage{amsmath}
\usepackage{amssymb}
\usepackage{bbm}

\usepackage{tikz}
\usetikzlibrary{positioning,calc}
\usetikzlibrary{arrows.meta}

\usepackage{pgfplots}
\usetikzlibrary{decorations.pathreplacing,matrix,positioning,calc}

\usepackage{lscape} 							
\usepackage{lipsum} 							
\usepackage{booktabs} 						
\usepackage{amsmath,amsfonts,amssymb} 	
\usepackage{array}							
\usepackage{textcomp}						
\usepackage{exscale} 							
\usepackage{tabularx} 						
\usepackage[natbibapa]{apacite}
\usepackage{ragged2e} 						
\usepackage{multirow}           					
\usepackage{eurosym} 						
\usepackage{verbatim}						
\usepackage{float}							
\usepackage{longtable}
\usepackage[toc,page]{appendix}
\usepackage{pdfpages}
\usepackage{rotating}
\usepackage{caption}
\usepackage{adjustbox}
\usepackage{longtable,lscape}
\usepackage{smartdiagram}
\usepackage[flushleft]{threeparttable}
\usepackage{numprint} 
\usepackage{siunitx}
\usepackage{tabularx}
\usepackage{numprint}
\usepackage{adjustbox}
\usepackage{placeins}

\newcommand{\co}[1]{{\color{red}\textbf{[}#1\textbf{]}}}
\newcommand{\ind}{\perp\!\!\!\!\perp}

\definecolor{darkgreen}{rgb}{0,0.5,0}

\usepackage{booktabs, caption, makecell}

\usepackage{threeparttable}

\usepackage{makecell}

\definecolor{plotblack}{named}{black}
\definecolor{plotgrey}{gray}{0.5}
\definecolor{plotblue}{HTML}{273A8F}
\definecolor{plotgreen}{HTML}{2C5C34}
\definecolor{plotred}{HTML}{B33D3D}
\definecolor{plotorange}{HTML}{DE7500}

\usepackage [autostyle, english = american]{csquotes} 

\usepackage[all]{nowidow} 

\title{{\Large \textbf{How to deal with machine learning bias in economic history\thanks{We want to thank Alexandra de Pleijt for thoughtful comments. We further want to thank participants at the ``Big Data and Machine Learning in Economic History'' workshop at the Uppsala History of Inequality and Labor Lab.  All code to reproduce results included in this manuscript is available on \url{github.com/juliuskoschnick/Machine-learning-and-bias-in-economic-history}}
}}\\
}

\author{Torben S. D. Johansen, Julius Koschnick, Christian Vedel (University of Southern Denmark)\footnote{University of Southern Denmark, Economics Department, Campusvej 55, Odense, Denmark}\\}

\begin{document}

\maketitle

\thispagestyle{empty}

\vspace{4ex}


\vspace{6mm}

\begin{abstract}		
    Machine learning (ML) has rapidly transformed economic history, lowering costs of digitization, data linkage, and imputation, and making information in historical text usable at scale. This paper offers a practical guide to using these tools well. 
    However, ML tools have also created new problems. Prediction errors are often systematically correlated with covariates of interest, so even highly accurate models can distort and sometimes reverse coefficients, and standard validation cannot detect this. Given that ML tools often perform worse for historical data, this problem is especially severe for the field of economic history. We also identify a solution to this problem. We show that recent debiasing methods can correct such bias for a wide class of applications, using a small, randomly sampled set of expert-coded labels while retaining the efficiency of large-scale prediction. We organize the field with a taxonomy of three ML tasks, survey the literature along it, and indicate where debiasing applies and where validation against proxies remains the only recourse. We close with best-practice guidance on digitization, model choice, and reproducibility.
	\vspace{12pt} \newline
	\thispagestyle{empty} \noindent \textbf{Keywords}: \textsc{Machine learning, Large language models, bias correction, economic history} \newline
\end{abstract}
\newpage

\setcounter{page}{1}

\section{Introduction}

Machine learning (ML) tools have radically shifted the research possibility frontier in the economic history and economic growth literature. New tools have made it easier to digitize data, create proxies for missing variables, link data, and capture information stored in text. This paper provides an overview over recent advances in the field.

A key problem is that these new methods introduce new forms of bias. Often, we treat ML-generated data as if they were observed data rather than predictions. This is conceptually incorrect. By construction, ML models produce predictions about an unobserved ground truth. This means that predictions necessarily contain prediction errors that can give rise to bias. Here, validating the data is not enough. Models with high accuracy might still have large structural bias. Recent literature has shown that such bias can be severely amplified in downstream applications \citep{egami2024using}. Therefore, this paper makes the key argument that the discipline does not live up to its own statistical standards when treating ML predictions as observed data. 

The problem is especially relevant for economic history, because ML tools tend to perform worse with historical sources than with the modern data they are trained on. For example, off-the-shelf large language models (LLMs) project modern categories onto the past and struggle with archaic language and period rhetoric \citep{bender2021,fang2024bias,CraneKarraSoto2025TotalRecall}. 
We illustrate this with two empirical exercises, a BERT classifier that loses accuracy as text is rewritten into older styles, and a frontier LLM that diverges from careful reading of seventeenth-century sources. Because these errors correlate with period, language, and complexity, they are systematic and can easily lead to bias in downstream results. 



Yet, recent statistical advances offer the toolset to actively address bias from ML models in economic history. Luckily, for a surprisingly large range of cases, debiasing frameworks \citep{angelopoulos2023prediction, egami2023using, egami2024using, ludwig2025large, carlson2025unifying} can fully address the problem. In this approach, expert historians annotate a random subsample of observations to construct a ``gold-standard'' reference. This data is then used to estimate the bias of ML predictions and correct for it in downstream applications. The debiased coefficients are consistently estimated while preserving the efficiency gains and statistical power afforded by large-scale ML predictions.

We argue that this approach is important for producing credible estimates based on ML predictions for historical data. Crucially, the debiasing approach combines the efficiency of ML tools with the expertise of the careful historian.  Moreover, we argue that any careful construction of a gold-standard should rely on careful source criticism and historians' close reading techniques. 
We illustrate this framework with an empirical example where we estimate the gender-wage gap in 1920 Stockholm. We demonstrate how ML predictions can lead to severe bias and how debiasing can recover the true estimate.

Next, the paper provides guidance on when debiasing can be applied in economic history. For this, we organize ML use in economic history into a taxonomy of three types of tasks, \textit{ML to replace human annotators}, \textit{ML to account for missing data}, and \textit{ML for new measures}. The taxonomy directly leads to decision rules for debiasing: Debiasing can always be applied to the first type, where in all cases, a gold standard can be constructed. It applies to the second whenever the missing values can be validated within an observed source. Finally, it does not apply to the third, where measures are fundamentally new and beyond human annotation capabilities, so that researchers must instead validate against proxies and placebo tests. We survey the literature along this taxonomy and argue that a large share of applications fall into the first category, so that debiasing covers a large share of ML work in the field. 

Our recommendation is important for the field of economic history, as ML tools have been widely and successfully applied in the discipline. Their use has generally shifted the research frontier and expanded the size of feasibly labeled datasets enormously \citep{korinek2023generative}.\footnote{Recent work shows that for many modern-day tasks, ML tools match or even exceed expert human annotators \citep{gilardi2023chatgpt,tornberg2024large,Asirvatham2026,Stewart_2026}.}  Just as important, ML makes the information stored in historical text usable at scale, providing direct evidence on beliefs, norms, political attitudes, scientific knowledge, and institutional practice that structured data capture only imperfectly. Recent work uses these tools to label historical sources \citep{dell2023american}, link individuals across historical and administrative records \citep{bjerregaard2023cohort, torbenpaper}, classify occupations \citep{dahl2026}, assign subject classes to publications \citep{koschnick2025feedback,Koschnick2025}, and extract sentiment, emotion, and thematic structure from textual and visual material \citep{gentzkow2019measuring,posch_raz_2025_doux_commerce,lagakos2025american,voth2024image,Gorin2025stateoftheart}. \citet{hufe2026chronos} present a framework for working with LLMs in historical sources. Moreover, History LLMs (HLLMs), large language models pre-trained on historical data with an explicit knowledge cutoff may help us to better understand path dependency and contingency and may offer insights into the values and psychological traits of historical societies \citep{goettlichetal2025,varnum2024large}.\footnote{\cite{MacBERTh_2021} and \cite{manjavacas-arevalo-fonteyn-2022-non} have base-trained historical BERT models, while \cite{goettlichetal2025} have created the first LLM chatbot with a knowledge cutoff before the First World
War. Other historical BERT models include several multilingual releases from the Bavarian State Library (Bayerische Staatsbibliothek). \cite{underwood2025} describe a small HLLM (774m parameters) trained on text from 1880--1914; to the best of our knowledge, the model has not yet been released.}

Using these tools credibly, however, requires new standards, above all a debiasing framework. With this, the paper complements recent overviews of ML in economics \citep{gentzkow2019text,dell2025deep,ludwig2025large,ash2023text,beach2022historical} and economic history \citep{Grajzl2025,grajzl2023machine}, as well as practical guides to operating these tools \citep{korinek2023generative,combes2022review,ferrara2026practitioners}. Relative to these, this paper focuses on the problem of statistical bias when applying ML tools and provides extensive guidance on how to address potential bias.

The paper proceeds as follows. Section~\ref{sec: debiasing} introduces the debiasing framework formally and develops the Stockholm gender-wage example in full. Section~\ref{sec:model_performance} documents why ML performance degrades on historical data. Section~\ref{sec:taxonomy} develops the taxonomy and, in Sub-Sections~\ref{subsec:ml_to_replace_human_annotators}--\ref{subsec:ml_for_new_measures}, treats each task in turn, surveying the literature and indicating where debiasing applies and where validation and placebo tests are the only recourse. Section~\ref{sec:best_practice} offers best-practice guidance on model choice, digitization, and reproducibility. Section~\ref{sec:conclusion} concludes.

\section{Debiasing}
\label{sec: debiasing}

\subsection{Motivating example: Estimating the gender wage gap in 1920 Stockholm}
\label{sec: debiasing-example}

ML tools are now broadly available to economic historians.
However, even accurate models can still yield severely biased predictions. We illustrate this with a classifier that correctly predicts gender 81 percent of the time yet attenuates the estimated gender wage gap by 0.109 log points, about 11 percent of the true gap.

We use individual income data for 3,271 taxpayers in Stockholm in 1920 from \cite{bengtson2024}, which record names, incomes, and true genders.\footnote{The \cite{bengtson2024} register assigns gender based on available archival evidence; we treat these assignments as our benchmark. In some applications, researchers would not have access to true gender and would need to infer it from names and/or other information.}
For the sake of the example, we assume that gender is not labeled but can be predicted with a machine learning model. 
Our machine learning method is a Naive Bayes model trained on 2,000 Scandinavian name-gender pairs.\footnote{Appendix~\ref{sec:nb_classifier} describes the model and training data.}
The model is deliberately simple and achieves a precision rate of 97.6\% and an F1 score of 79.6\%.\footnote{It achieves 97.6\% precision but only 67.2\% recall: the model rarely labels a man as a woman, but misclassifies 32.8\% of women as men, resulting in biased predictions.}

We proceed with treating the model's predictions as if they represented true genders and regress log wages on the predicted gender dummy. 
The result is a gender wage gap of $-0.713$ log-points; however, as we will show, this estimate is biased.
To get a sense of the bias, we compare this to what we label the ``oracle'' estimate---an estimate using the true value of gender from the original data.
This places the gap at $-0.822$ log-points, suggesting our
naïve ML estimate is biased by 0.109 log points. 
The size of the bias is large: the literature on the gender wage gap is typically concerned with decomposing total gender wage gaps of a similar magnitude \citep{kleven2019, goldin2014}. The estimate is confident and wrong.

\begin{figure}[ht]
    \centering
    \includegraphics[width=1\textwidth]{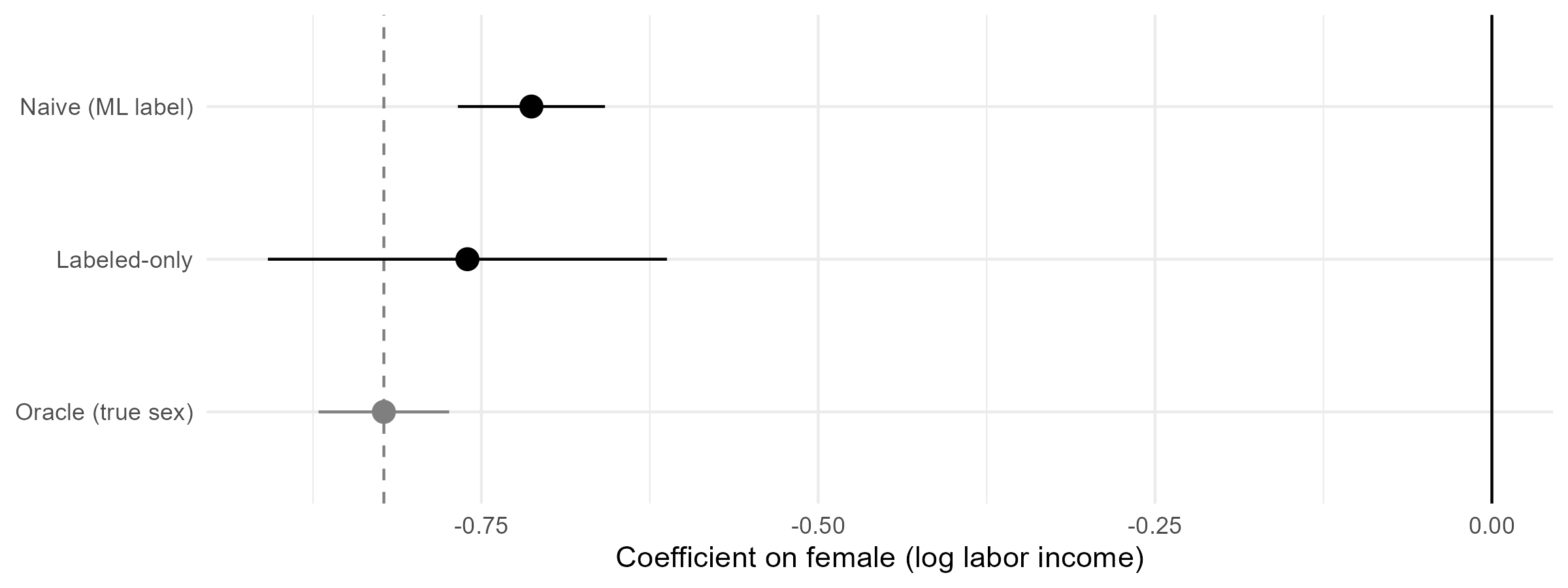}
    \caption{Gender wage gap estimates, Stockholm 1920}
    \label{fig:stockholm_coefplot-v2}
    \begin{minipage}{1\linewidth}
        \vspace{5pt}
        \footnotesize
        \textit{Notes:}
        The figure shows the results of regressing log income on (1) true gender of all individuals in our sample, (2) inferred gender based on ML predictions using names, and (3) true gender for a 10\% subsample.
        Oracle refers to the estimate using true gender for the full sample.
        Naïve ML refers to the estimate using inferred gender for the full sample.
        Labeled-only refers to the estimate using true gender for a 10\% subsample.
        Horizontal bars show 95\% confidence intervals.
        The vertical, dashed line shows the ``true'' gender wage gap (as proxied by the oracle estimate).
        The vertical, solid line shows the case of no gender wage gap.
        Data: \cite{bengtson2024}.
    \end{minipage}
\end{figure}

A natural solution is to get back to the original source and obtain true genders.
For the example's sake, suppose a researcher has a budget/time constraint and can code up to 10\% of the sample; in our case, this constitutes 326 individuals sampled at random.\footnote{Rather than sampling 10\% of the 3,271 observations, which would result in 327 labeled cases, we independently sampled every observation with 10\% probability, which explains the one observation discrepancy. This ensures independent sampling, which is a prerequisite for our later debiasing approach. In practice, this makes a minimal difference.}
We can now run our analysis on this subset to obtain an unbiased estimate of the gender wage gap. Because we discard 90\% of the observations, however, our confidence interval is much wider than the confidence intervals of the oracle estimator or the naïve ML approach.
Figure~\ref{fig:stockholm_coefplot-v2} shows the point estimates and confidence intervals for all three approaches.

Regardless of whether we pursue a naïve ML approach or use only the 10\% labeled subset, our results are sub-optimal.
In the case of naïvely using ML predictions, the fact that we obtain biased results invalidates the entire purpose of the analysis.
And while we obtain unbiased results when using only labeled data, these are measured with significant noise due to the smaller sample size.
However, rather than rely on either approach, we can combine them: ML predictions supply precision by exploiting the full sample, and the labeled subsample supplies the correction for systematic errors.

\FloatBarrier
\subsection{Debiasing ML Predictions}
\label{sec: debiasing-ml-preds}

Debiased ML seeks to remedy the problem of biased ML models by exploiting access to ``gold-standard'' labels for a subset of ML predictions.
At their core, these methods---most prominently prediction-powered inference \citep[PPI;][]{angelopoulos2023prediction} and design-based supervised learning \citep[DSL;][]{egami2023using, egami2024using}---address the problem by estimating and subsequently addressing the bias of ML systems on a subset of observations for which ``gold standard'' labels are available.\footnote{Earlier approaches to correcting for ML-predicted labels in downstream analyses include \citet{wang2020methods} and \citet{fong2021machine}. PPI and DSL require weaker assumptions about the relationship between ML predictions and true values, allowing them to act as nearly universal debiasing methods.}

To use debiased ML, one needs access to (1) an unlabeled corpora of interest, (2) an ML model, such as an LLM, that can label these, and (3) a random subset of the corpora with ``gold standard'' labels.
At a high level, the first step of the procedure is then to use the ML model to create labels for \textit{all} samples in the corpora.
The second step is to compare these predictions with ``gold standard'' labels for the subset where these are available.
This allows one to characterize the bias in the ML predictions.
The third step is then to estimate the quantity of interest---such as a mean, quantile, or regression coefficient---using the ML predictions but accounting for the bias estimated in the second step.
In terms of our previous example, the idea is to use the 10\% subsample with observed genders to correct the bias of our ML model that inferred genders for the full sample.

To illustrate how debiasing operates in the simplest case, consider a researcher interested in the average wage who wishes to use PPI. This differs from our running example, where the prediction (gender) enters as a regressor rather than as the outcome being averaged. 
This researcher has access to a large corpus containing the relevant information, but producing ``gold standard'' labels would be prohibitively expensive.
Denote by $N$ the total number of samples in the corpus and by $n$ the number of ``gold standard'' labels the researcher is willing to create, with $n \ll N$.\footnote{For convenience, and without loss of generality, we assume the data is ordered such that labels are available for observations $1$ through $n$.}
Let $X_i$ denote the text for sample $i$ and $Y_i$ the wage.
The researcher always observes $X_i$ but only observes $Y_i$ for the subsample of ``gold standard'' labels she created.
She further has access to an ML system that produces predictions of $Y_i$ given $X_i$; call this $f$.
She can then estimate the mean $\theta$ of $Y_i$ in the following ways:
\begin{align*}
    \hat{\theta}_{\text{GS}} &= \frac{1}{n} \sum_{i = 1}^nY_i & \text{Using only ``gold standard'' labels}
    \\
    \hat{\theta}_{\text{ML}} &= \frac{1}{N} \sum_{i = 1}^N f(X_i) & \text{Using ML predictions naïvely}
    \\
    \hat{\theta}_{\text{DB}} &= \frac{1}{N - n} \sum_{i = n + 1}^N f(X_i) - \underbrace{\frac{1}{n} \sum_{i = 1}^n \left( f(X_i) - Y_i \right)}_{\text{Bias correction term}} & \text{Using debiasing (PPI)},
\end{align*}
where $\hat{\theta}_{\text{GS}}$ denotes the sample average using only manually created ``gold standard'' labels, $\hat{\theta}_{\text{ML}}$ denotes the average using only ML predictions, and $\hat{\theta}_{\text{DB}}$ denotes the average computed using debiasing.
She is interested in which of these estimates is best.

Consider first $\hat{\theta}_{\text{GS}}$.
Since the researcher has drawn a random sample, this will clearly provide an unbiased estimate of $\theta$, but since $n \ll N$, it might have high variance.
Consider next $\hat{\theta}_{\text{ML}}$.
If $f$ produces perfect predictions for $Y_i$, this is desirable compared to $\hat{\theta}_{\text{GS}}$ (since the variance will be lower), but our ML system might produce biased predictions, in which case this estimate is meaningless.
Finally, consider $\hat{\theta}_{\text{DB}}$.
This uses ML predictions to exploit all $N$ samples while still remaining unbiased, due to the bias-correction term.
Bias-correction often helps one achieve higher efficiency compared to using only ``gold standard'' labels while maintaining unbiased estimates.
For large enough $N$ compared to $n$, debiasing will lead to a more precise estimate whenever the variance of residuals, $f(X_i) - Y_i$, is smaller than the variance of $Y_i$.
For additional details, and to see how DSL can be used in this setting, see Appendix~\ref{sec: ppi-vs-dsl}.

While more involved, this core idea also carries over to settings where the missing variable is an independent variable in a regression, such as is the case with the missing genders from the example in Section~\ref{sec: debiasing-example}; Appendix~\ref{sec: ppi-vs-dsl-rhs} shows how to use PPI and DSL in such settings.
Additionally, debiasing extends to a broad range of estimands beyond means and slopes.
Both PPI and DSL support linear, logistic, and Poisson regressions; DSL additionally handles high-dimensional fixed effects \citep{egami2024using}.
More recent advances extend debiasing to instrumental variables, difference-in-differences, and regression discontinuity designs \citep{carlson2025unifying, ludwig2025large}.\footnote{Software: The \texttt{ppi-python} Python package implements PPI \citep{angelopoulos2023prediction} and the \texttt{dsl} R package implements DSL \citep{egami2023using}. For IV, DiD, and RD extensions, see the replication archive accompanying \citet{carlson2025unifying}.}

These methods apply to any black-box ML model: when the gold-standard sampling probabilities are known by design, the bias correction has expectation zero regardless of the predictive accuracy of the ML model, meaning unbiasedness depends on the sampling design rather than the ML model. 
Precision gains, however, scale with the model's predictive ability, and when model performance is low, using only labeled data may provide more precise estimates.
Sampling may be uniform or stratified, but convenience sampling or ad hoc estimated sampling probabilities invalidate the bias correction and standard errors.

\subsection{Revisiting Our Example}

\begin{figure}[ht]
    \centering
    \includegraphics[width=1\textwidth]{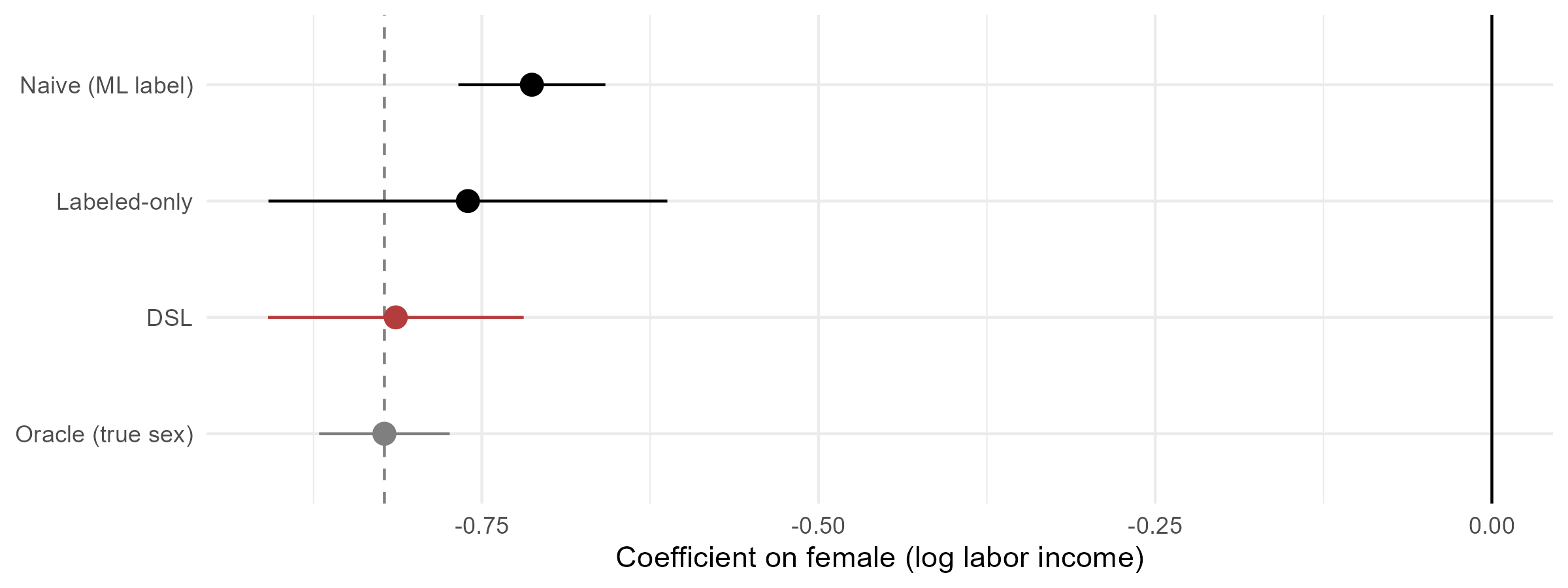}
    \caption{Gender wage gap estimates, Stockholm 1920}
    \label{fig:stockholm_dsl_coefplot-v2}
    \begin{minipage}{1\linewidth}
        \vspace{5pt}
        \footnotesize
        \textit{Notes:}
        The figure shows the results of regressing log income on (1) true gender of all individuals in our sample, (2) inferred gender based on ML predictions using names, (3) true gender for a 10\% subsample, and (4) inferred gender based on ML predictions using name, alongside DSL for debiasing.
        Oracle refers to the estimate using true gender for the full sample.
        Naïve ML refers to the estimate using inferred gender for the full sample.
        Labeled-only refers to the estimate using true gender for a 10\% subsample.
        DSL refers to the estimate using inferred gender for the full sample combined with DSL for debiasing.
        Horizontal bars show 95\% confidence intervals.
        The vertical, dashed line shows the ``true'' gender wage gap (as proxied by the oracle estimate).
        The vertical, solid line shows the case of no gender wage gap.
        Data: \cite{bengtson2024}.
    \end{minipage}
\end{figure}

We now use the random sample drawn at a 10\% probability to debias the ML estimates of the gender-wage gap for our Stockholm example using DSL.\footnote{The 10\% subsample consists of 326 observations, the same subsample used for the labeled-only estimate.} 
Figure~\ref{fig:stockholm_dsl_coefplot-v2} reports the debiased estimate as well as the three previous estimates (naïve ML, labeled-only, and the oracle), alongside 95\% confidence intervals.
The DSL estimate yields a coefficient of -0.814 [-0.909, -0.719], statistically indistinguishable from the oracle ($-0.82$ log-points), with a confidence interval 36\% narrower than the labeled-only estimate.
As the figure shows, debiasing delivers an estimate that is both unbiased and more precise than the labeled-only approach. 
It recovers consistent estimates of the true coefficients while preserving the efficiency and statistical power afforded by large-scale ML predictions.

\FloatBarrier
\section{ML performance with historical data}
\label{sec:model_performance}

The debiasing framework outlined above generalizes  to any use of ML predictions in economic analyses where it is possible to acquire gold standard data. Economic history, however, has particular reason to rely on it. The historical sources at the heart of the discipline often diverge sharply from the modern data on which today's models are trained. This gap makes prediction errors both larger and more systematic than in present-day applications. This section sets out why ML tools are often likely to misrepresent historical data and illustrates the argument with two empirical exercises.

The root of the problem is that most ML tools are pre-trained models---including BERT-style classifiers and all LLMs---and that historical data has a different structure than the modern corpora on which they were trained. For example, the expression of seventeenth-century sentiment is likely to differ from sentiment expressed on Twitter/X, which usually forms the backbone of modern sentiment classifiers. Likewise, historical place names or occupational descriptions are likely to differ from their modern counterparts. 

This mismatch between training target and application target gives rise to two structural problems. First, because LLMs are trained predominantly on text from recent decades, they project modern notions onto the past, giving rise to anachronisms and temporal contamination. Careful prompting, instructing a model to adopt the perspective of a period or group, mitigates this only partially. The model imitates a historical perspective using modern knowledge, and its knowledge of history degrades with temporal distance, leading it to blend historical and modern information when information runs out \citep{CraneKarraSoto2025TotalRecall}. Moreover, note that the historical texts that form part of LLM training data are themselves selected. They reflect survival bias and the overrepresentation of Western material in archives and libraries, and models may read sources literally, without regard to the political and societal position of their authors.

Second, historical text is syntactically and semantically harder, especially outside English, where training data is thin, and sources may refer to events or debates of which the model knows little. Both factors do more than lower average performance. They generate errors that are correlated with factors such as period, language, region, social context, or textual complexity that can cause bias in downstream settings.

One way to address the mismatch between training and target data is to use HistoryLLMs, LLMs pre-trained on historical data with clear knowledge cutoffs \citep{varnum2024large,goettlichetal2025}. Here, the knowledge cutoff guarantees the absence of contamination from modern data. However, given that these models are only trained on historical data, they are generally less capable than LLMs trained on modern data. Here, HistoryLLMs are likely to underperform on specifically challenging tasks. Moreover, just like modern LLMs, they are likely to reproduce societal biases from their training data. Finally, researchers might find it difficult to find HistoryLLMs that fit their specific time period. Hence, the problem of bias persists for uses of ML, and especially LLMs, in economic history.

\begin{figure}[ht]
    \centering
    \includegraphics[width=0.75\textwidth]{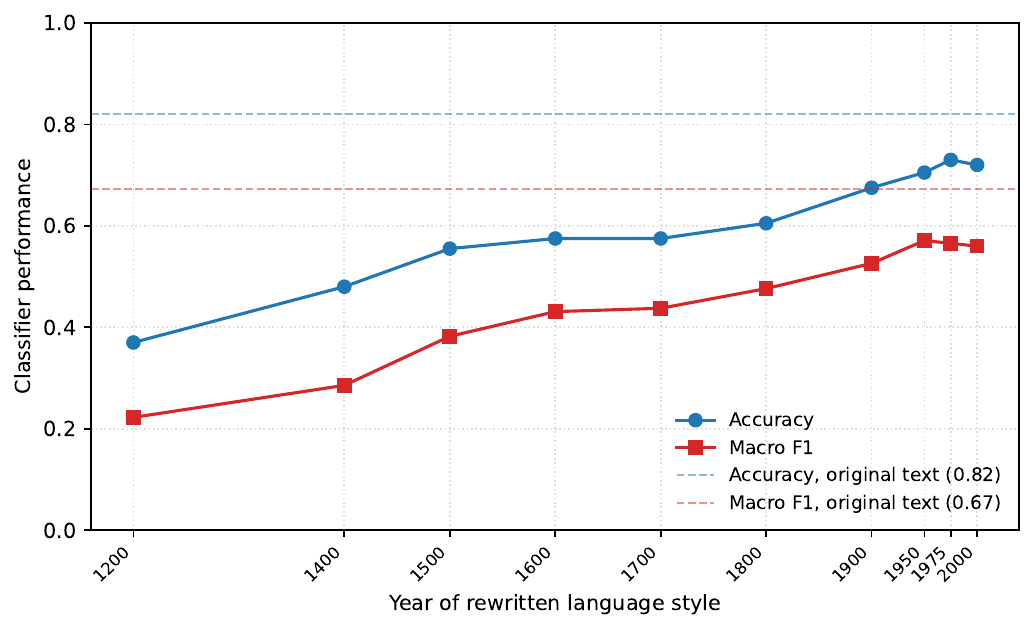}
    \caption{BERT model performance in classifying emotions by time period of linguistic style}
    \label{fig:performance-over-time}
    \begin{minipage}{1\linewidth}
        \vspace{5pt}
        \footnotesize
        \textit{Notes:} The figure plots the performance of a pre-trained BERT emotion
classifier (\texttt{j-hartmann/emotion-english-distilroberta-base}) in recovering the emotion of 200 texts randomly drawn
from the \texttt{dair-ai/emotion} dataset (with the corresponding labels of \textit{joy, sadness, anger, fear,
love}, and \textit{surprise}). Each text is rewritten by a \texttt{gpt-5.4-mini} LLM into
the written English style of the indicated time period.
Accuracy and macro-averaged F1 are computed for the same 200 observations at each period. Dashed lines report the classifier's performance on the original, un-rewritten (modern) texts.
    \end{minipage}
\end{figure}

\begin{figure}[ht]
    \centering
    \includegraphics[width=0.75\textwidth]{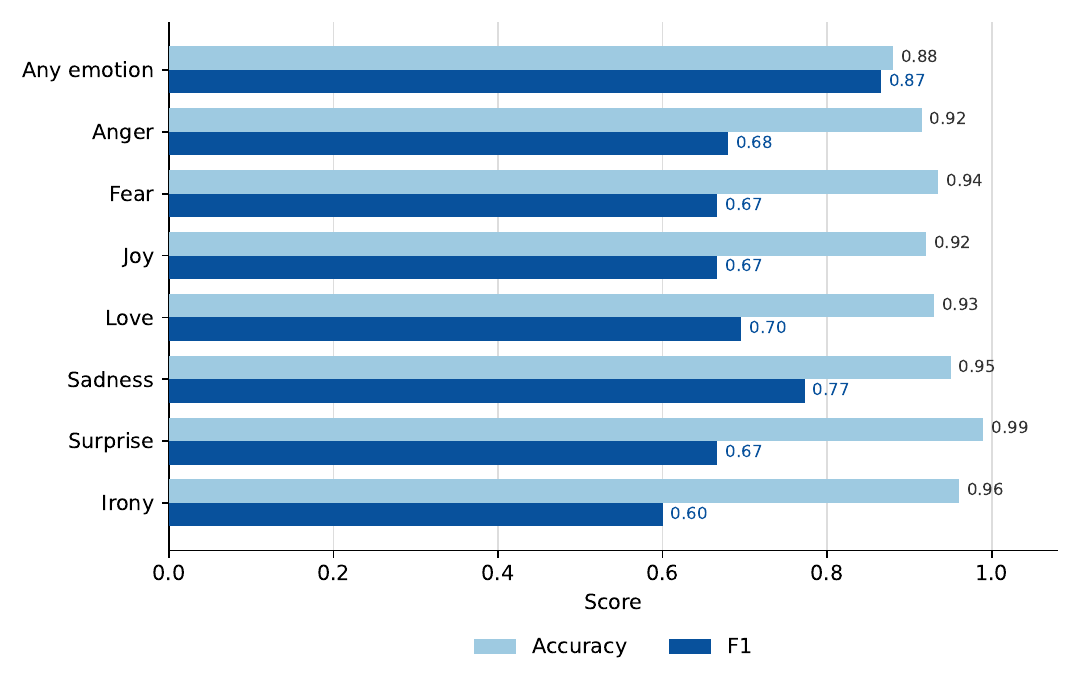}
    \caption{GPT-5.5 performance with real seventeenth century text data}
    \label{fig:llm-historical-test}
    \begin{minipage}{1\linewidth}
        \vspace{5pt}
        \footnotesize
    \textit{Notes:} The figure compares emotion and irony labels produced by a modern
large language model (\texttt{gpt-5.5}) against hand-coded labels for a random
sample of paragraphs drawn from seventeenth-century English books (EEBO).
For each category we report the share of paragraphs classified correctly
(accuracy) and the macro-averaged F1 score, treating the hand-coded labels as
ground truth ($N=200$ coded paragraphs). The model is given the original
early-modern text. While accuracy is uniformly high, F1 is substantially lower
across all categories and lowest for irony.
    \end{minipage}
\end{figure}

We illustrate the problem of performance degradation of LLMs when applied to historical text with a very simple quantitative exercise. We take the \texttt{dair-ai/emotion} dataset, one of the standard emotion evaluation datasets from the \textit{Hugging Face} platform, and ask an LLM (\texttt{gpt-5.4-mini}) to rewrite the texts in the linguistic style of different time periods. The \texttt{dair-ai/emotion} dataset comes with human-evaluated labels for the presence of the emotions of \textit{anger, fear, joy, love, sadness,} and \textit{surprise} in the text. We then use a state-of-the-art BERT classifier, \texttt{j-hartmann/emotion-english-distilroberta-base}, to classify the original and rewritten texts.\footnote{The model was chosen because it, crucially, was not trained on the \texttt{dair-ai/emotion} dataset.} We choose the BERT model as it strongly outperforms current LLMs on the evaluation dataset.\footnote{The BERT classifier reaches an accuracy of 0.82 and an F1 statistic of 0.67 in comparison to an accuracy of 0.59 for a GPT-5.4 model and an F1 statistic of 0.50 (in comparison: For a GPT-5.4-mini model, we get an accuracy of 0.59 and an F1 statistic of 0.47).}

Figure~\ref{fig:performance-over-time} shows that the classifier's performance strongly degrades the more archaic the underlying text becomes. Note that the LLM-rewritten text is most likely more comprehensible to present-day models than real historical text. As such, the exercise provides a conservative baseline. Yet, even in this simple case, we see a dramatic decrease in model performance when using archaic language.

Next, in a second exercise, we illustrate that even latest-generation LLMs can fail to accurately label historical data. For this, we draw a random sample of 200 paragraphs from Early English Books Online, a collection of full-text volumes from seventeenth-century England.\footnote{Concretely, we draw from the A28 sample from the Text Creation Partnership batch that contains Civil War and post-Civil War volumes.} Then, in a procedure similar to the previous exercise, we ask a \texttt{gpt-5.5} LLM to assign the emotions of \textit{anger, fear, joy, love, sadness, surprise}, and \textit{irony}. We add the irony component as it captures an integral part of seventeenth century rhetoric. The paragraphs were then independently labeled by the authors.

Figure~\ref{fig:llm-historical-test} compares the LLM-labels with the author labels. Although the LLM attains high accuracy (a macro-average of 0.94 across the six basic emotions), its F1 score is low---0.69 overall and just 0.60 for irony--- revealing substantial disagreement that the accuracy figures conceal.\footnote{For a quick outline of accuracy and f1, see our practioner's guide in appendix section~\ref{app:sec:validation_statistics}.} We take this as evidence that even last-generation LLMs often diverge from human labels on challenging historical text. That said, note that we do not have ex-ante proof that the author-labels are superior the LLM labels. To prove our point that LLMs can make significant mistakes at such tasks, Appendix~\ref{app:sec:close_reading} employs the methods of the critical historian. We conduct a source criticism and close reading exercise for two of these paragraphs. We place them in historical context and interpret the concrete rhetorical devices employed. We show that the LLM misses historical context, misinterprets the structure of the source and author standpoints, and struggles with seventeenth century rhetorical devices. Our close-reading section thus underlines the need for close-reading and interpreting text against the broader source-specific context, even when using last-generation LLMs.

Altogether, the two exercises make the same point from different directions. The emotion exercise shows that a strong classifier's accuracy falls steadily as the text becomes more archaic, even though LLM-rewritten prose is almost certainly easier for modern models than genuine historical sources. The Early English Books exercise shows that even a latest-generation LLM diverges from a careful human reading on seventeenth-century material, and the close reading in Appendix~\ref{app:sec:close_reading} shows that these are not random slips but failures to grasp historical context, source structure, and period rhetoric. This degradation is a substantial concern for the field, because structural errors of this kind can easily be amplified in downstream tasks. Our gender-wage example in Section~\ref{sec: debiasing-example} already showed how biased predictions of gender from names distort a downstream regression, and \cite{egami2024using} show that even models with precision above 90 or 95 percent can produce bias large enough to flip the sign of a coefficient. The exercises above suggest that historical applications routinely fall below these thresholds. We therefore recommend that researchers in economic history who use ML predictions apply the debiasing framework whenever possible. The next section provides further guidance on when this is possible by developing a taxonomy of ML tasks in economic history.

\FloatBarrier
\section{A taxonomy of ML tasks in economic history}
\label{sec:taxonomy}

Having established that ML predictions on historical data are prone to systematic bias, we now ask when that bias can be corrected. The debiasing framework of Section~\ref{sec: debiasing} is our preferred recommendation, yet it cannot be applied everywhere: It requires a gold-standard sample, and for some tasks it is impossible to construct ground truth data. To clarify where the framework applies and where researchers must instead fall back on validation and placebo tests, we develop a taxonomy of three types of ML tasks in economic history. For each type, we survey how the literature has used these tools and recommend how to address the resulting bias.

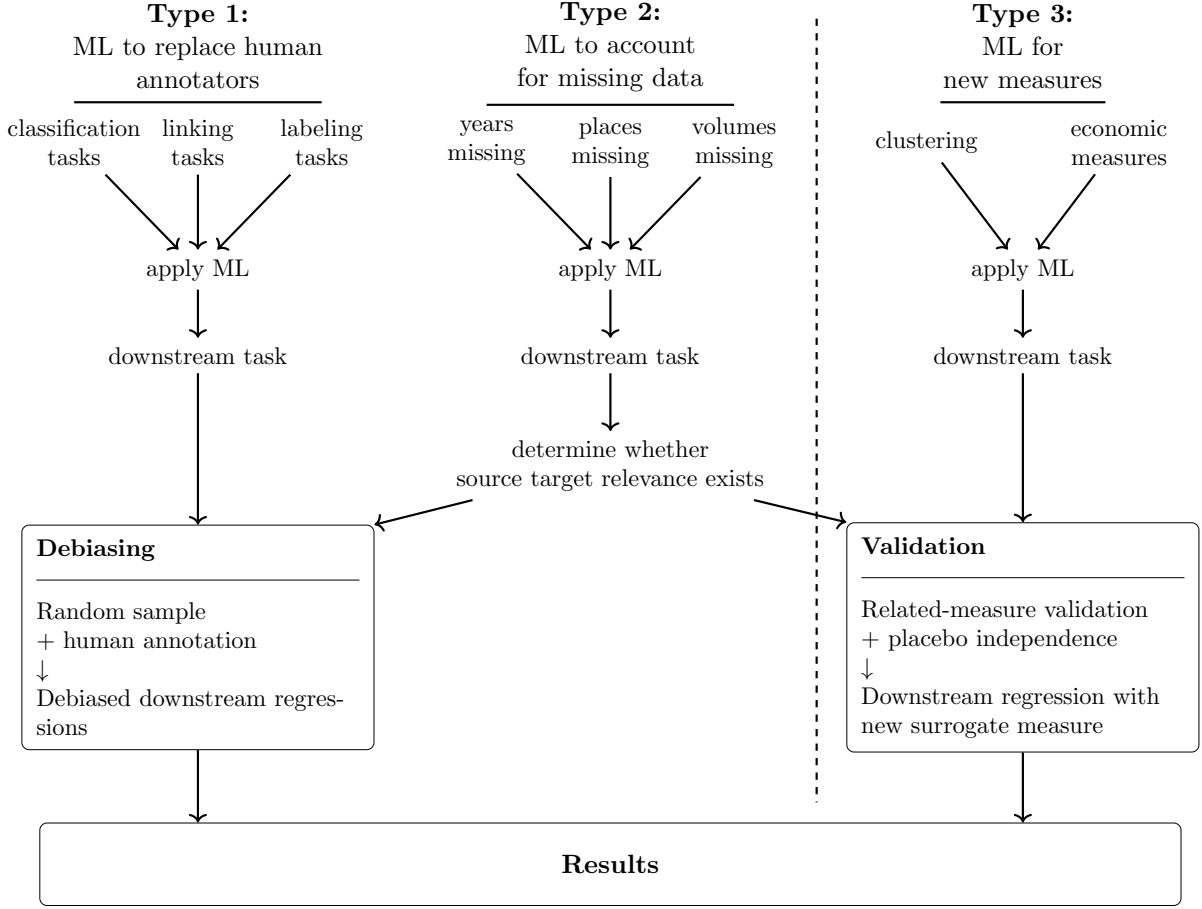
\begin{figure}[!ht]
    \centering
    \begin{adjustbox}{width=1\textwidth}
        \begin{tikzpicture}[
  font=\normalsize,
  arr/.style={->, line width=0.9pt},
  hdr/.style={align=center},
  small/.style={font=\small, align=center},
  mid/.style={font=\small, align=center},
  box/.style={
    draw,
    rounded corners=3pt,
    align=left,
    inner sep=6pt,
    text width=4.7cm,
    font=\small
  },
  resultsbox/.style={
    draw,
    rounded corners=3pt,
    align=center,
    inner sep=10pt,
    minimum height=1.2cm,
    minimum width=16.6cm
  }
]

\def\xL{-6}
\def\xM{0}
\def\xR{6}

\node[hdr] (h1) at (\xL,0) {\textbf{Type 1:}\\ ML to replace human\\annotators};
\node[hdr] (h2) at (\xM,0) {\textbf{Type 2:}\\ ML to account\\for missing data};
\node[hdr] (h3) at (\xR,0) {\textbf{Type 3:}\\ML for \\new measures};

\draw[line width=0.9pt] (h1.south) ++(-1.8cm,-2pt) -- ++(3.6cm,0);
\draw[line width=0.9pt] (h2.south) ++(-1.8cm,-2pt) -- ++(3.6cm,0);
\draw[line width=0.9pt] (h3.south) ++(-1.2cm,-2pt) -- ++(2.4cm,0);

\draw[dashed, line width=0.9pt] (3,0.4) -- (3,-11.0);

\node[small] (c1a) at (\xL-1.8,-1.4) {classification\\tasks};
\node[small] (c1b) at (\xL,-1.4) {linking\\tasks};
\node[small] (c1c) at (\xL+1.8,-1.4) {labeling\\tasks};

\node[small] (c2a) at (\xM-1.8,-1.4) {years\\missing};
\node[small] (c2b) at (\xM,-1.4) {places\\missing};
\node[small] (c2c) at (\xM+1.8,-1.4) {volumes\\missing};

\node[small] (c3a) at (\xR-1.4,-1.4) {clustering};
\node[small] (c3b) at (\xR+1.4,-1.4) {economic\\measures};

\node[mid] (ml1) at (\xL,-3.25) {apply ML};
\node[mid] (ds1) at (\xL,-4.5) {downstream task};

\node[mid] (ml2) at (\xM,-3.25) {apply ML};
\node[mid] (ds2) at (\xM,-4.5) {downstream task};

\node[mid] (ml3) at (\xR,-3.25) {apply ML};
\node[mid] (ds3) at (\xR,-4.5) {downstream task};

\node[mid] (rel2) at (\xM,-6.1) {determine whether\\source target relevance exists};

\node[box] (b1) at (\xL,-8.6) {
\textbf{Debiasing}\\[-2pt]
\rule{\linewidth}{0.4pt}\\[4pt]
Random sample\\
+ human annotation\\
$\downarrow$\\
Debiased downstream regressions
};

\node[box] (b3) at (\xR,-8.6) {
\textbf{Validation}\\[-2pt]
\rule{\linewidth}{0.4pt}\\[4pt]
Related-measure validation\\
+ placebo independence\\
$\downarrow$\\
Downstream regression with new surrogate measure
};
\node[resultsbox] (RES) at (0,-11.9) {\textbf{Results}};

\draw[arr] (c1a) -- (ml1);
\draw[arr] (c1b) -- (ml1);
\draw[arr] (c1c) -- (ml1);

\draw[arr] (c2a) -- (ml2);
\draw[arr] (c2b) -- (ml2);
\draw[arr] (c2c) -- (ml2);

\draw[arr] (c3a) -- (ml3);
\draw[arr] (c3b) -- (ml3);

\draw[arr] (ml1) -- (ds1);
\draw[arr] (ml2) -- (ds2);
\draw[arr] (ml3) -- (ds3);

\draw[arr] (ds1) -- (b1.north);
\draw[arr] (ds3) -- (b3.north);

\draw[arr] (ds2) -- (rel2);
\draw[arr] (rel2) -- (b1.north east);
\draw[arr] (rel2) -- (b3.north west);

\coordinate (t1) at (RES.north -| b1);
\coordinate (t3) at (RES.north -| b3);

\draw[arr] (b1.south) -- (t1);
\draw[arr] (b3.south) -- (t3);

\end{tikzpicture}
    \end{adjustbox}
    \caption{Bias and solutions in different ML tasks in economic history}
    \label{fig:three-tasks}
    \begin{minipage}{1\linewidth}
        \vspace{5pt}
        \footnotesize
        \textit{Notes:}
        The graph shows our separation of tasks in economic history into three distinctive types: \textit{ML to replace human annotators}, \textit{ML to account for missing data}, and \textit{ML for new measures}. 
        For each type it lists tasks that fall within that type of ML application. We then illustrate how bias can be addressed in downstream tasks. For type 1, \textit{ML to replace human annotators}, where it is possible to humanly annotate a random sample, we recommend a debiasing approach. For type 3, \textit{ML for new measures}, where random annotation is usually impossible, we recommend that researchers rely on a validation and placebo approach instead. For type 2, \textit{ML to account for missing data}, researchers have to determine whether they can turn their task into either a type 1 or type 3 problem by determining whether source-target relevance exists. Please refer to the main text for a full explanation.
    \end{minipage}  
\end{figure}

Figure~\ref{fig:three-tasks} provides an overview of our taxonomy of ML tasks in economic history. The taxonomy follows the nature of the task to which ML is applied. The three tasks are
Type 1: \textit{ML to replace human annotators}, in which ML helps speed up the work a human would otherwise perform, such as transcription; Type 2: \textit{ML to account for missing data}, in which ML helps predict the value of genuinely missing data, such as helping to fill in measures for years where source material is unavailable; and Type 3: \textit{ML for new measures}, in which ML is used to construct entirely new (often surrogate) measures of economic variables.

It turns out that debiasing is a viable solution to all ML applications within type 1 problems, \textit{ML to replace human annotators}. However, for type 3 problems, \textit{ML for new measures}, we lack gold-standard validation data, so that researchers have to rely on careful validation and placebo-testing. Finally, for type 2 problems, \textit{ML to account for missing data}, the structure of the missing data determines whether we can turn type 2 problems either into type 1 problems with debiasing as a viable solution or into type 3 problems, where debiasing techniques cannot be applied.

Below, we discuss all three types of tasks in detail. It turns out that type 1 problems cover a wide range of ML applications in the field of economic history. Hence, it appears that debiasing can be widely applied within the discipline and that it can significantly raise the credibility of estimations relying on ML predictions.

\FloatBarrier
\subsection{Task 1: ML to replace human annotators}
\label{subsec:ml_to_replace_human_annotators}

For Type 1 problems, ML is mainly labor augmenting, where the labor of human expert labelers is substituted for ML. This substitution makes it possible to scale projects that previously heavily relied on human annotators, e.g., for classification, linking, or labeling tasks, to a much larger scale. As a result, it allows economic historians to study new datasets and questions that were previously practically infeasible, such as assigning topics to millions of historical newspaper and newswire articles \citep{dell2023american,Silcock2024}. Yet, using ML tools also runs the risk of introducing systematic bias into downstream estimates.

Recent applications of ML as a substitute for human annotation in economic history span several types of tasks. The most common body of work extracts latent attributes, such as topics or political sentiment, from historical documents.

First, recent research has shed light on the development of science, religion, and culture. \cite{almelhem2023enlightenment} use bag-of-words
methods on English book titles from 1500 to 1900 to document the separation
of science from religion and a rising belief in progress. In a similar topic modeling approach for 17\textsuperscript{th} century England,
\cite{grajzl2024macroscope} find evidence that innovations in religion stimulated focus on science, but find little evidence of secularization. \cite{Curtis2026seeds} use $k$-means for identifying topics in the publications of European scholars between 1350 and 1800. Finally, \cite{gatti2025text} analyze Michelangelo's letters using a combination of bag-of-words methods and BERT models to trace the relationship between Michelangelo's emotions and external life factors.

Next, a broad range of papers have adopted ML tools to analyze political opinion in history based on large text corpora. \cite{Giommoni2026} use an LLM to quantify the political sentiment of parliamentary speeches during the French Revolution. They validate its output against human annotators and report accuracies between 0.677 and 0.952. In a similar vein, \cite{dePleijt2026} apply an LLM to infer political sentiment toward Parliament from students' full-text publications during the English Civil War. Likewise, \cite{gentzkow2019measuring} measure the development of political partisanship from U.S. congressional speeches from 1873 to 2016 using a plug-in estimator.  

Moreover, unsupervised methods have also been deployed to identify technological change \citep{liu2025technology} and construct life fulfillment indicators from American life histories \citep{lagakos2025american}.
ML tools have also been applied to visual information from photographs and paintings: \cite{voth2024image} measure visual style in U.S. high-school yearbook photographs from 1930 to 2010, documenting a long-run convergence of male and female styles followed by renewed stylistic polarization after the 1970s. \cite{Gorin2025stateoftheart} extract emotional expressions from images spanning world history from 1400 onward, opening up the systematic study of affect at a large historical scale.

\cite{dahl2026} introduce \textit{OccCANINE}, a custom trained model for assigning HISCO codes to occupational strings that has been applied widely in the literature. Finally, \cite{liu2025technology} use a \textit{gpt-4o-search-preview} model to classify historical US census occupations into \textit{cognitive}, \textit{manual}, or \textit{interpersonal} categories. 

A final group of contributions applies ML to record linkage and entity recognition at large scale.
\cite{feigenbaum2016} develops a supervised ML approach for linking individuals across historical US censuses, while \cite{price2021} use genealogical links as training data to construct linked full-count census panels.
More recently, \cite{arora2024} introduce a multimodal contrastive-learning model that links Japanese firms across noisy historical documents.
In a related entity-resolution application, \cite{Silcock2024} use deep-learning models to deduplicate 2.7~million unique US newswire articles and disambiguate the individuals mentioned in them.
Other recent contributions use machine learning to link US census data and vital records, expand on the use of genealogical links as training data, examine the construction of training data for supervised linkage, and develop an ML approach for linking Chinese historical records \citep{bailey2023creation,buckles2025breakthroughs, feigenbaum2025examining, yu2026machine}.

A common denominator of these applications of ML in economic history is the reconstruction of descriptive data from secondary or indirect sources, such as record linkages, entity recognition, the assignment of classification systems, or the extraction of political sentiment from text data. This introduces the need to use tools for the construction of the desired economic measures, e.g., using ML tools. 

In comparison, a researcher working on contemporary questions might decide to directly collect this data, for example using administrative identifiers for linkage or running surveys to quantify political opinion. Yet, economic historians cannot create new data on the past that was not collected before. Therefore, ML predictions are necessary for a large set of questions in economic history where the set of direct observables is small. 

However, it is crucial that, for downstream tasks, the prediction error rate is less important than the presence of potential structural biases in the model output. Were prediction errors classical measurement error, their consequences would be predictable and depend on the role of the predicted variable. In practice, ML errors are rarely classical: even small structural errors can lead to severe bias in any direction in downstream tasks \citep{egami2024using}. Therefore, even in cases of low average prediction errors, we still recommend that researchers adopt a debiasing framework.

In the age of ML, economic historians need to be careful historians. The careful historian should start with careful source criticism \citep{ranke_1824_kritik,bloch_1954_historian,tosh_2015_pursuit}. This process involves, at the very least, placing the source in temporal and spatial historical context. Furthermore, the role of the author of the source, their intentions, and their situation within the historical society should be critically discussed. Next, the careful historian should document the layers of how the text and data was collected or composed. Here, it should be asked whether there are gaps or whether the data-creating process differed across different parts of the source. Researchers might also want to verify the contents of the source against alternative sources with similar content.

Only after these first stages of source criticism should the coding of gold standard labels be begun. When using third parties to code labels, it should be ensured that they have sufficient knowledge of the historical background and are familiar with the particular context of the source at hand. 

We further encourage researchers to make both their source criticism and the information provided to expert annotators as explicit as possible. Ideally, the documentation should include a brief source critique, an overview of the annotators’ relevant background knowledge, and a complete account of the coding rules specified in the codebook. In this way, the adoption of the debiasing framework will also increase the prominence and transparency of source criticism and engagement with historical context in research in papers within the discipline.

\subsection{Task 2: ML for missing data}

The second class of tasks uses ML not to substitute for a human annotator but to recover data for which we lack source material. Here, data might have been lost, never recorded, or omitted. We call these \textit{Type~2} problems, and they differ from \textit{Type~1} problems in one fundamental respect: There is no extant record from which a human coder could, even in principle, read off the true value. The information is genuinely missing, and the task is therefore one of imputation rather than transcription.

We organize these problems around two objects: a \textit{source}, in which the variable of interest is observed, and a \textit{target}, in which it is missing. An ML model learns the mapping from features to outcome on the source and applies it to the target.\footnote{In some applications the source is used only to validate or debias predictions, rather than to train the predictive model itself.}

The defining difficulty of Type~2 problems is that direct validation is unavailable. In a Type~1 problem the researcher can always construct a gold standard, because the truth is recoverable from the source. In a Type~2 problem the truth in the target is, by construction, unobservable, so predictions cannot be checked against ground truth precisely where it matters. Our central recommendation is to ask whether this difficulty can be circumvented by turning the Type~2 problem into a Type~1 problem. In a surprising number of cases it can, and the route runs through the source. Although the outcome is missing in the target, it is observed in the source; if the researcher is willing to treat the source as informative about the target---an assumption we label \textit{source--target relevance}---then predictions can be validated within the source, and debiasing becomes available. This assumption is not structurally different from historical practice in everyday archival work, where the relevance of a source to the question at hand is a routine judgment.

To make this precise, suppose the source and target are drawn from the same population, with the outcome $Y_i$ observed only in the source. Let $R_i \in \{0, 1\}$ indicate membership in the source ($R_i = 1$) versus the target ($R_i = 0$), and let $X_i$ denote the features used for prediction---names, in the motivating example of Section~\ref{sec: debiasing-example}. The researcher then observes the single dataset $(X_i, R_i \cdot Y_i)$, where the term $R_i \cdot Y_i$ masks the outcome whenever an observation belongs to the target. Source observations can be used to validate predictions whenever
\begin{equation}
    (X_i, Y_i) \ind R_i,
\end{equation}
that is, whenever the joint distribution of features and outcome is independent of whether the outcome is observed. This corresponds to Assumption~2 of \cite{carlson2025unifying},\footnote{Note that you can also have a version of this assumption conditional on covariates.} and it is what licenses carrying the estimated $X_i$--$Y_i$ relationship from the source to the target. A natural validation strategy under this assumption is cross-validation within the source. Researchers can train on part of the source, evaluate on the held-out remainder, and iterate. This recovers exactly the ingredients a Type~1 problem requires: predictions paired with verified labels, so the debiasing methods of Section~\ref{sec: debiasing} apply directly. A version of the assumption that conditions on covariates is also available \citep{carlson2025unifying}, and likewise supports debiasing.\footnote{As in any debiasing application, the same observations must not be used both to fit the predictive model and to estimate its bias; cross-fitting or a held-out validation set avoids the resulting leakage.}

This yields a simple decision rule. If the researcher can validate even a small share of predictions against observed outcomes---directly, or indirectly through source--target relevance---the problem reduces to Type~1 and can be debiased. If no such validation is possible, the problem instead resembles a Type~3 problem, for which only indirect validation against proxies is available.

One caution applies throughout. Source--target relevance is an assumption about selection into observation, not a remedy for a biased source. If the model is trained on a source whose labels are themselves systematically biased and then applied to the target, that bias propagates to the predictions and cannot be undone by debiasing; it must be confronted directly, as in the motivating example from Section~\ref{sec: debiasing}. The assumptions above discipline how the source generalizes to the target; they do not sanitize the source.

The literature spans the full range, from problems that reduce cleanly to Type~1 to problems that remain stubbornly Type~3---recall that Type~2 problems resolve into one or the other. The clearest cases come from a growing body of work that uses names as indicators of cultural outcomes. \cite{Abramitzkyetal2020} construct a name-based measure of assimilation, capturing how relatively likely a name is to be foreign rather than domestic; \cite{bentzen2024assimilate} adapt the approach to measure whether Danish migrants retained a Danish name-identity; \cite{knudsen2024those} uses the prevalence of common names as an indicator of collectivist culture; and \cite{Kok2026} uses names to identify a Jewish identity in Dutch historical data. The approach traces to the Black Name Index of \cite{fryer2004bni}. These measures share a common structure and are all closely related to a Naive Bayes classifier \citep{IdiotBayes2001}, the same method used in the application above: each takes a source in which names are paired with an outcome---names and ethnic identity in \cite{Abramitzkyetal2020}, for instance---and applies the learned mapping to a target in which only names are observed. Because the measure is at bottom a prediction, its accuracy can be checked on a held-out sample of the source, and where source--target relevance is plausible, debiasing can be implemented. These are therefore natural Type~1 candidates.

A second group of applications addresses missing occupational outcomes, and here the appropriate classification depends on the variable. Much of the literature on occupational scoring can be seen in this light, as can \cite{Paker2025PredictingPast}, who recovers missing occupational data from an auxiliary source. \cite{Sobek1995} introduced occupational scores in the IPUMS census data. The idea is to use the 1950 U.S. Census, where both occupations and wages are observed, as the \textit{source}, and then apply the resulting occupation--wage mapping to \textit{target} settings where occupations are observed but income is not. \cite{Inwood2019candiantrouble} show how the source--target relevance assumption can be problematic in this setting. \cite{Saavedra2020mloccscore} introduces a machine-learning method for constructing occupational scores from U.S. Census data, addressing some of these concerns by including covariates. For a concrete, verifiable quantity such as wages, source-based validation is feasible, and the problem can often be treated as \textit{Type~1}. For an abstract construct such as occupational status, where no ground truth can be pinned down, it is not, and the problem is better treated as \textit{Type~3}. 

\cite{koch2024augmenting} mark the limit of what validation can deliver. They estimate historical GDP per capita from biographical data on notable individuals, training an elastic-net model on country- and region-period cells where GDP is observed and predicting it for cells where it is not. In our terms, the observed cells form the source and the missing cells the target. The testable claim---that biographical data predict GDP within the source---is not the binding one; the binding assumption is that this relationship travels to the target cells, and that cannot be tested directly. The authors accordingly validate against external proxies for economic output, including urbanization, height, well-being, and church-building activity. Such proxy validation strengthens the credibility of the relevance assumption but does not discharge it, which is why the exercise is best understood as a Type~3 problem.

In short, Type~2 tasks pose an essential question. Can the source--target relationship be validated where it matters? Where the missing outcome is verifiable and source--target relevance holds, the problem reduces to Type~1 and the debiasing framework of Section~\ref{sec: debiasing} applies. Where the target is an abstract construct with no recoverable ground truth, it becomes a Type~3 problem. We turn to those problems next, where direct validation is unavailable by construction and credibility must instead rest on proxies and placebo tests.

\FloatBarrier
\subsection{Task 3: ML for new measures}
\label{subsec:ml_for_new_measures}

The third type of task is different in kind. Rather than recover a value that a source records and a human could in principle verify, ML can be used to construct entirely new measures. Examples include continuous measures of innovation, sentiment, or political stance based on text data. While these create fundamentally new knowledge, they are also the most challenging for credible inference. Because these measures go beyond anything previously observed, there is no ground truth against which they could be checked. Hence, the debiasing framework, which relies on a gold-standard sample, cannot be applied. The task of this section is twofold: It first provides an overview of how new measures are constructed and have transformed the existing literature. It then proposes a validation-placebo strategy for how fundamentally new measures can be made credible in the absence of a gold standard.

Generally, many cases of \textit{ML for new measures} have been built on information from the embedding spaces of transformer models. Embedding spaces are a core component of architectures such as BERT and modern large language models and provide a high-dimensional geometric representation of learned semantic relationships (768 dimensions per token in BERT-base).\footnote{There is strong evidence that embedding spaces encode a wide range of semantic properties such as human associations between words \citep{pennington2014glove, caliskan2017}. BERT and other LLMs further provide representations of contextual semantics \citep{ethayarajh2019contextual}. Furthermore, another strain of research suggests that higher BERT layers correspond more closely to semantics while lower layers correspond more strongly to syntactical understanding \citep{reif2019visualizing}. } 
If a researcher is interested in the relationships among texts within a large corpus, then she can map the texts into an embedding space where they are represented as individual vectors. The geometric location of these vectors is informative about the semantic relationships among texts.

\begin{figure}[ht]
    \centering
    \begin{subfigure}[b]{0.48\textwidth}
        \centering
\begin{tikzpicture}[
    >=Stealth,
    arith/.style = {-{Stealth[length=7pt]}, line width=1.1pt, black}
  ]

  \useasboundingbox (-0.3,-1.0) rectangle (4.9,3.7);

  \coordinate (O)  at (0,0);
  \coordinate (k)  at (3.4,1.0);   
  \coordinate (km) at (2.2,0.05);  
  \coordinate (R)  at (2.8,2.3);   
  \coordinate (q)  at (3.05,2.2);  

  \draw[gray!50, -{Stealth[length=5pt]}] (O) -- (4.3,0);
  \draw[gray!50, -{Stealth[length=5pt]}] (O) -- (0,3.6);

  \draw[gray!60, dotted, line width=0.9pt] (O) -- (R);

  \draw[arith]         (O)  -- (k);
  \draw[arith, dashed] (k)  -- (km);
  \draw[arith]         (km) -- (R);

  \fill[black] (k) circle (1.8pt);
  \fill[black] (R) circle (1.8pt);
  \draw[black, line width=0.9pt] (q) circle (2.3pt);

  \node[font=\footnotesize\itshape, right=2pt] at (k)                              {king};
  \node[font=\footnotesize, anchor=west]       at ($(k)!0.5!(km) + (0.12,-0.02)$)  {$-\,\textit{male}$};
  \node[font=\footnotesize, anchor=west]       at ($(km)!0.7!(R) + (-0.12,0)$)    {$+\,\textit{female}$};
  \node[font=\footnotesize\itshape, right=2pt] at (q)                              {queen};

  \node[anchor=west, font=\small] at (0.05,3.35)
        {$\textit{king}-\textit{male}+\textit{female}\;\approx\;\textit{queen}$};

\end{tikzpicture}
        \caption{Arithmetic on word meanings}
        \label{fig:kingqueen}
    \end{subfigure}
    \hfill
    \begin{subfigure}[b]{0.48\textwidth}
        \centering
\begin{tikzpicture}[
    >=Stealth,
    docvec/.style = {-{Stealth[length=7pt]}, line width=1.1pt, black},
    lab/.style    = {font=\footnotesize, align=left}
  ]

  \useasboundingbox (-0.3,-1.0) rectangle (4.9,3.7);

  \coordinate (O) at (0,0);

  \draw[gray!50, -{Stealth[length=5pt]}] (O) -- (4.1,0);
  \draw[gray!50, -{Stealth[length=5pt]}] (O) -- (0,3.7);

  \coordinate (p1) at (30:3.4);
  \coordinate (p2) at (62:3.0);
  \draw[docvec]         (O) -- (p1);
  \draw[docvec, dashed] (O) -- (p2);

  \draw[black, line width=0.8pt] (30:1.05) arc (30:62:1.05);
  \node[font=\small] at (46:1.45) {$\theta$};

  \node[lab, anchor=west]       at (p1) {Scientific\\paper 1};
  \node[lab, anchor=south west] at (p2) {Scientific\\paper 2};

  \node[anchor=west, font=\small] at (-0.2,-0.6)
        {$\cos\theta$: similarity of two papers};

\end{tikzpicture}
        \caption{Similarity between documents}
        \label{fig:papers_cosine}
    \end{subfigure}
    \caption{Geometry of an embedding space}
    \label{fig:embedding_geometry}
    \begin{minipage}{1\linewidth}
        \vspace{5pt}
        \footnotesize
        \textit{Notes:}
        Both panels are stylized two-dimensional projections of a high-dimensional embedding space. In contrast, real spaces have hundreds of dimensions (768 per token in BERT-base). Panel~\subref{fig:kingqueen}: Starting from the embedding of \textit{king}, subtracting \textit{male} and adding \textit{female} (tip to tail) lands close to the embedding of \textit{queen}---arithmetic on the vectors mirrors a relationship between the words' meanings. Panel~\subref{fig:papers_cosine}: The cosine of the angle $\theta$ between two document vectors measures their semantic similarity, close to one when aligned and close to zero when unrelated. The same geometry that supports word arithmetic also lets researchers build continuous similarity measures over longer texts.
    \end{minipage}
\end{figure}
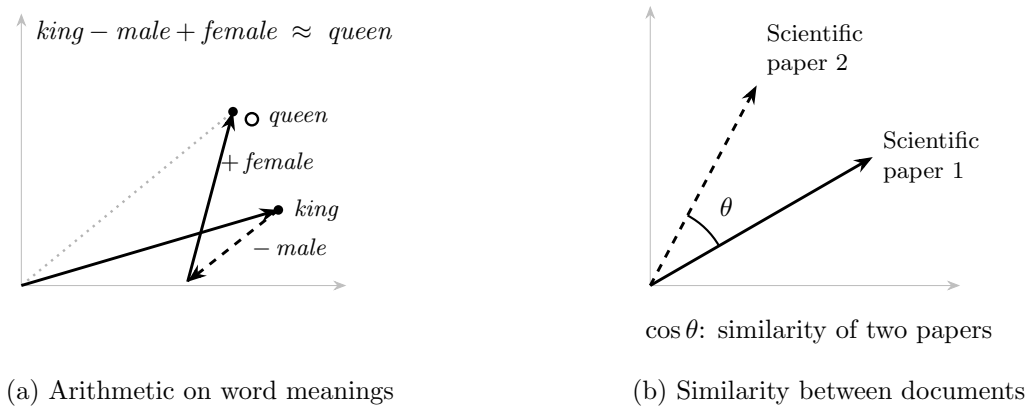

The idea underlying embedding-space measures builds on a long tradition in linguistics and information retrieval in which meaning is inferred from patterns of use. The distributional view of language holds that words with similar meanings appear in similar contexts \citep{harris1954distributional,firth1957synopsis}. Early computational implementations represented documents as vectors of weighted word frequencies and measured their similarity by the angle between vectors \citep{salton1975vector}, while latent semantic analysis used matrix decompositions to recover lower-dimensional semantic structure from term-document co-occurrence patterns \citep{deerwester1990indexing}. Modern language models operate according to a related principle: Pieces of text are represented as vectors in a high-dimensional space, and these representations allow the model to predict the next token with high accuracy. Many type~3 measures exploit this same geometry for a different purpose. Rather than using embeddings only to generate or classify text, researchers use distances and directions in embedding space as empirical measures of semantic relationships.
A standard illustration of embedding spaces is given by mapping the words \textit{queen}, \textit{king}, \textit{female}, and \textit{male} (Figure~\ref{fig:embedding_geometry}, Panel~\subref{fig:kingqueen}). Within an embedding space, subtracting the vector for \textit{male} from \textit{king} and adding the vector for \textit{female} returns a vector very close to \textit{queen}.

To capture semantic similarity between texts within an embedding space, researchers usually use cosine similarities defined as:

\[
\text{cos}(\mathbf{x}_i, \mathbf{x}_j) 
= \frac{\mathbf{x}_i \cdot \mathbf{x}_j}{\|\mathbf{x}_i\| \, \|\mathbf{x}_j\|} 
= \frac{\sum_{k=1}^{d} x_{ik} x_{jk}}{\sqrt{\sum_{k=1}^{d} x_{ik}^2} \, \sqrt{\sum_{k=1}^{d} x_{jk}^2}},
\]
where \(\mathbf{x}_i, \mathbf{x}_j \in \mathbb{R}^d\) denote the embedding vectors of two texts and \(d\) is the dimensionality of the embedding space. $\text{cos}(\mathbf{x}_i, \mathbf{x}_j)$ then measures the angle between $\mathbf{x}_i$ and $\mathbf{x}_j$ (Figure~\ref{fig:embedding_geometry}, Panel~\subref{fig:papers_cosine}). Semantically aligned texts yield values close to one and unrelated texts yield values close to zero.\footnote{Note that the use of embedding spaces in BERT models is often associated with a range of technical challenges. During learning, embedding spaces can get compressed and exhibit anisotropy, leading to unrelated words inhabiting parts of the same vector space \citep[see, e.g.,][]{gao2019representation}. Likewise, BERT models trained on classification objectives exhibit poor performance on similarity prediction tasks. Here, base layers should be fine-tuned using, e.g., \textit{simple contrastive learning of sentence embeddings} approaches (SimCSE) \citep{gao2021simcse}; for a historical application see \cite{koschnick2025feedback}.}

Hence, cosine similarities and other vector operations make it possible to capture the relationship between different vectors of text, for example, the semantic similarity between two scientific papers (Figure~\ref{fig:embedding_geometry}, Panel~\subref{fig:papers_cosine}). Moreover, they can be used to create completely new proxies for economic concepts of interest such as innovation, sentiment, or political alignment based on text data alone. Given the extremely large amount of information processed for these comparisons, constructing such measures would have been intrinsically impossible for human annotators. The following paragraphs discuss two literatures that were transformed by the arrival of embedding-space measures: the study of innovation and (political) sentiment in economic history.

First, the literature on innovation has long struggled with the problem that we cannot measure innovation directly. Instead, the literature always had to rely on proxies such as patent citations. Although patent citations have been widely and successfully applied in the literature on innovation \citep{griliches1998patent,trajtenberg1990penny,Jaffe1993}, they only constitute imperfect proxies. It is well known that patent citations only explain a small share of patent value \citep{trajtenberg1990penny,gambardella2008value}, capture a selected subsample \citep{griliches1998patent,moser2005patent}, and reflect non-innovation factors, such as social networks, legal norms, or the structure of a field \citep{roach2013lens}. Therefore, a burgeoning literature in economics \citep{kelly2021measuring} and science of science  \citep{park2023papers} has used information from text data as an input for alternative measures of innovation. Text-based innovation measures employed in economic history include \cite{Iaria2018}, \cite{kelly2021measuring}, \cite{Chiopris2024}, \cite{LaMelaFrankemolleTell2025}, \cite{GRAJZL2026}, and \cite{billington2026success}.

The most influential approach here is \cite{kelly2021measuring}, who base their measure on the simple logic that patents are more innovative if they are closer to future rather than previous work. They operationalize this using a bag-of-words \textit{tf-idf} approach and apply it to U.S. patents between 1840 and 2010. However, as we argue in Section~\ref{subsec:bag_of_words}, bag-of-words approaches are always information inefficient in comparison to LLM-based approaches and suffer from semantic bias. Therefore, \cite{Ganguli2024patenttext} argue in favor of using embedding-space measures instead of bag-of-words methods for capturing measures of patent similarity. Following this, \cite{Aghion2023} and \cite{boeing2024anatomy} apply embedding-based innovation measures to modern-day settings. As a further step forward, \cite{koschnick2025feedback} applies the innovation index logic from \cite{kelly2021measuring} to a BERT model and further generalizes the approach to spillover measures. Moreover, \cite{koschnick2025feedback} demonstrates how the underlying embedding space can be aligned with historical time periods by using historically pre-trained and corpus-fine-tuned BERT models. It is then used to test Joel Mokyr's feedback loop hypothesis. Given that embedding-space measures are much more capable of handling non-standardized texts, these approaches have the potential to allow us to uncover dynamics of innovation wherever historical text is available, including at a local level, across different genres of text, and at high frequency. 
Therefore, embedding-space innovation measures can significantly broaden the study of innovation in economic history.

Second, the literature on (political) sentiment analysis has also been transformed by new embedding-space measures. Traditionally, sentiment was captured through lexical approaches based on the frequency of keywords that captured sentiment. However, such measures would usually miss both context and complex formulations, leading to relatively poor performance. The arrival of transformer models, such as BERT, then made it possible to construct context-sensitive sentiment measures reaching an accuracy on established datasets of up to $\sim$90\% \citep{devlin2019bert,ash2023text,alahmadi2025generalizing}. Here, researchers either calculate embedding-space similarities between keywords and the full target text or they construct emotion-specific classifiers.\footnote{For keyword and embedding-space approaches, researchers have also found it helpful to scale their measures by distance to multiple sets of words capturing opposing sentiment or political opinions \citep{ash2021}.} Because the outcomes are continuous measures of sentiment based on large-scale comparisons across text, they go beyond the capabilities of human annotators and create fundamentally new information. This approach has made it possible to study diverse questions of human beliefs, attitudes, and sentiment in the past. Applications include quantifying 200 years of newspaper sentiment \citep{van2024almost}, local trust in US newspapers \citep{posch_raz_2025_doux_commerce}, and gender stereotypes \citep{ash2024gender}.

Yet, for these types of Type~3 problems, it is intrinsically hard to account for potential ML bias. Because all measures create fundamentally new information, it is intrinsically impossible to construct a gold standard. Researchers might still worry about ML bias from, e.g., anachronistic representations or from the difficulty of disentangling meaning from style.\footnote{It seems to the authors that this problem is not only practical but inherent: as style can evoke emotion, irony, and nuance, it necessarily intersects with semantics.}
Our recommendation is to borrow the logic of \textit{surrogates}, long used in medicine \citep[see, e.g.,][]{prentice1989surrogate,pepe2003statistical} and recently formalized for economics by \cite{athey2025surrogate}. Surrogates are substitutes for a concept or variable that cannot be measured at the current moment. They are widely used in economics, for example, years of schooling for human capital, test scores for cognitive skills, patent citations for innovation, or height for nutrition and disease environments in history.\footnote{As an example from economic history, take \cite{izdebski2020pollen} who use pollen data to infer agricultural activity in ancient Greece. Naturally there is no direct validation available. However, they are able to validate their pollen measure against archaeological field surveys, shipwrecks, and evidence on oil and wine presses from other areas.} We argue that new economic measures based on ML predictions are methodologically similar to the case of surrogates and should follow similar methodological standards.

Concretely, we propose that researchers should establish two conditions: First, a strong correlation between the new surrogate measure and ideally multiple relevant proxies, and second, the absence of systematic correlation with a set of conceptually similar but economically unrelated measures.

The first condition builds on recent work on surrogate measures in economics by \cite{athey2025surrogate}. The central argument is that surrogates can only be made credible indirectly by showcasing their correlation with another established proxy. Moreover, agreement with a single proxy can be fragile. Just as conditioning on additional covariates makes the unconfoundedness assumption more plausible, consistency with multiple, qualitatively distinct proxies makes it more likely that surrogates behave similarly to the unobserved variable of interest.

The second condition is that the measure shows no systematic correlation with conceptually similar but economically unrelated quantities. This placebo condition guards against the characteristic failure of a surrogate identified by \cite{athey2025surrogate}, which economists will recognize as Goodhart's law. A surrogate can move without the underlying construct moving with it. For example, clickbait can increase user interaction without leading to any purchases. Likewise, text-based surrogates might respond to changes in genre, length, or stylistic convention rather than to the variable of interest they are supposed to capture. Demonstrating that the surrogate is unrelated to placebo treatments acts as an active guard against these behaviors.

\cite{koschnick2025feedback} illustrates the application of the validation-placebo approach. Its new innovation index is shown to correlate with both patent citations and the number of published editions while remaining unaffected by several placebo treatments. Additionally, an LLM-based rewriting of the corpus shows that the result is not an artifact of writing style. However, note that validation of this kind establishes a measure as a usable surrogate, but it does not make it identical to the proxy against which it is checked. Text-based measures and conventional proxies capture different processes and may legitimately diverge. For example, text-based measures of innovation are different from patent citations in the same way that patent citations are different from patent value.

Finally, this validation does not need to be repeated from scratch in every study. \cite{athey2025surrogate} propose a shared ``library'' of surrogate indices whose validity has been established across applications, so that a measure vetted once can be reused with confidence. Economic history is unusually well suited to such an effort, since a surrogate can be checked not only against multiple proxies but across periods and places. Such a library would be, for Type~3 problems, what the debiasing framework is for Types~1 and~2, namely the means by which ML-based measurement is held to the discipline's standards of credible inference.

\FloatBarrier
\section{Best Practice}
\label{sec:best_practice}

Having set out the framework for clean inference, we now turn to the practical choices that often arise when applying ML tools in economic history.
We first discuss best practices for implementing the debiasing framework. We then consider broader challenges in applications of ML to economic history, including model selection, the digitization of historical sources, and replicability, and offer concrete solutions to each.

\subsection{A practical guide to debiasing}

Applying the debiasing framework takes three steps: \textit{random sampling}, \textit{annotation}, and \textit{application} of the estimator. The order mirrors the workflow most researchers face, from validating an ML model's predictions to producing the final debiased estimate.

\paragraph{Step 1: Validation and random sampling}
When an ML prediction task of either type in Section~\ref{sec:taxonomy} is finished, the natural next step is to validate the predictions against human-verified data. This is good practice whether or not the downstream use-case will be debiased. The researcher draws a random sample, using a known sampling procedure, from the prediction data.\footnote{When ML is used to train a model, performance is usually validated against a held-out subset the model has not seen.} At a minimum, rigor requires reporting human-verified performance with accuracy, precision, recall, and/or F1, defined in Section~\ref{app:sec:validation_statistics}.

Debiasing assumes that the ground truth for each ML-predicted observation can be established given enough time and resources. The researcher draws a random subsample of $n$ observations from the ML-labeled data to serve as the gold standard.\footnote{As a reminder $n$ refers to the small sample of gold-standard labeled data, and $N$ refers to the full corpus.} How large should $n$ be? Because the efficiency of the debiased estimate rises with $n$, the answer depends on the statistical power each application needs. What drives $n$ is how much the predictions cut residual variance. The precision gain depends on the variance of the prediction error $f(X)-Y$ relative to the variance of $Y$. A practical route is to pilot with a small $n_0$, estimate the residual variance, and top up to the precision the application requires. In practice, values of 500 to 3,000 are common \citep{egami2024using,Rister2025,Yang_et_al_2025_LLM_annotation}, and \cite{Yang_et_al_2025_LLM_annotation} warn against fewer than 200 labels in complex data structures. Against a large corpus of $N$ predictions, hand-coding a few hundred to a few thousand observations is a small cost that still guarantees consistent downstream estimates.

A crucial assumption of the debiasing framework is that the sample follows a known design. Every observation needs a strictly positive, known probability of selection. The design need not be uniform, however. Subject to this requirement, researchers may over-sample observations that are hard to predict or where accuracy matters most, provided the sampling weights enter the estimator \citep{egami2024using}.

In practice, we recommend the following workflow: Assign each observation a known probability $\pi_i$, record it, draw the sample on those probabilities, and pass $\pi_i$ to the estimator as weights. Default to uniform $\pi_i$ (simple random sampling) unless there is a specific reason to stratify.

\FloatBarrier
\paragraph{Step 2: Annotation}

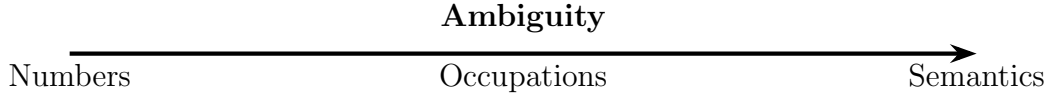
\begin{figure}[htp!]
    \centering
    \begin{tikzpicture}[
        >=Stealth,
        axis/.style   = {-{Stealth[length=10pt,width=7pt]}, line width=1.6pt},
        every node/.style = {font=\large},          
        label/.style = {anchor=north}               
      ]
        \draw[axis] (0,0) -- (12,0);
        
        \node[anchor=south,font=\large\bfseries] at (6,0.15) {Ambiguity};
        
        \node[label] at (0,0)  {Numbers};
        \node[label] at (6,0)  {Occupations};
        \node[label] at (12,0) {Semantics};
    
\end{tikzpicture}
    \caption{The ambiguity scale}
    \label{fig:ambiguity_scale}
\end{figure}

As a next step, expert coders assign the labels. In economic history, annotators need domain knowledge of the period at hand. This is the work of the careful historian, who, as set out in Section~\ref{sec:model_performance}, engages critically with the source and its context while hand-coding. Researchers should decide how much time the task warrants and what domain training research assistants need. Where the task is ambiguous, they can also check agreement between the ML predictions and several independent coders, as in \cite{lagakos2025american}.

The skill required for annotation rises with the semantic ambiguity of the data. Figure~\ref{fig:ambiguity_scale} arranges this ambiguity along a single scale. At the left end, the mapping from source to data is exact. Extracting numbers from tables may be hard for ML tools, but the output is unambiguous and trivial for a human to check. Toward the middle, conceptual ambiguity appears. Assigning occupations to HISCO codes, for example, requires judgment, because a profession can plausibly belong to more than one category. Coding it correctly demands careful reading of the relevant rule books, careful documentation of those rules and the coding decisions, and attention to how occupational terms shifted over time---an engagement with the historicity of the source.

At the right end, the mapping depends on semantics and context. Inferring historical topics or political opinion, for instance, requires annotators to navigate the full ambiguity of natural language alongside the context of the source. This demands a good knowledge of the period's debates and rhetorical conventions. It is here that the careful historian's role fully comes to bear.

\paragraph{Step 3: Application of the debiasing framework}

After annotating the dataset, researchers apply the debiasing framework, relying on existing software. The design-based supervised learning (DSL) estimator \citep{egami2023using} is available as the \texttt{dsl} R package \citep{dsl2025}, and prediction-powered inference (PPI) \citep{angelopoulos2023prediction} as the \texttt{ppi-python} package in Python. Moreover, the number of ready-to-use software implementations of debiasing frameworks is likely to increase in the imminent future. 

We further stress the versatility of the debiasing approach. It is possible to apply it to, e.g., models with fixed effects or difference-in-difference estimators. Moreover, it can also be used for data that is aggregated in downstream applications. As long as both the labeled data and the sampling probabilities are aggregated by the same function, inference remains consistent.

Debiasing is guaranteed to be consistent, and with a reasonably good prediction model and large enough $N$, it yields more precise estimates than the hand-coded sample alone.\footnote{Take, for example, our earlier LHS PPI case. Provided sufficiently large $N$, debiasing yields a more precise estimate whenever the variance of the residuals, $f(X_i) - Y_i$, is smaller than the variance of $Y_i$.} Section~\ref{sec: debiasing-example} shows both properties: Debiasing recovered a consistent estimate of the gender-wage gap, which the uncorrected ML predictions understated, and it supported meaningful inference, whereas the gold-standard sample alone remained severely underpowered.

Debiasing therefore lets researchers extend questions that require careful historical labeling to datasets far larger than manual coding of $n$ alone would permit. This scaling is what makes the approach valuable for economic history. We can trace political opinion not only across a few individuals' texts but across entire corpora of publications, judge the representativeness of a life history by comparing its key characteristics, such as education, occupation, and successes or failures, against large sets of relevant biographies; or trace attitudes toward violence, gender, justice, or religion across large corpora and over time \citep[for pioneering work, see][]{lagakos2025american}.

These applications can expand the scope of economic history. Comparing such large-$N$ analyses against the discipline's classical narratives can challenge received notions and push the field's knowledge frontier.

\FloatBarrier
\subsection{Thinking about the scale of your problem}
\label{subsec:scale_of_the_problem}
A programmer, the joke goes, will happily spend 200 hours automating a task that would take 10 hours by hand. Economic historians face the same gamble each time a new corpus invites automation. 
An important consideration here is the \emph{scale} of the problem. A practical way to gauge scale is to ask: \emph{How many hours would this take to do manually?} A precise estimate is rarely possible, but reasoning in magnitudes---tens, hundreds, or thousands of hours---is usually enough to narrow down which class of solution is worth considering.

There are three broad approaches to solving big data problems in economic history:

\begin{itemize}
    \item \textit{Mode 1 --- ``Type it yourself''}. Until a few years ago, this was the only option. Fixed costs are close to zero, but the marginal cost is high. This is often the right choice for genuinely small tasks, or where any automation would take more set-up than the work itself.
    \item \textit{Mode 2 --- ``Off-the-shelf solutions''}. Here, the fixed cost is higher (assembling a small pipeline), but marginal costs are lower. This could be an LLM to extract structured data from biographies, or one of the many specialized models on, e.g., \textit{HuggingFace}.
    \item \textit{Mode 3 --- ``Build your own solution''}. Custom systems carry the highest fixed costs (training, annotation, engineering, deployment) but very low marginal costs once running. This becomes attractive when $N$ is large enough that marginal costs dominate, and it puts large-scale data work within reach of even small research teams. See, e.g., \cite{torbentranscription2024}, who re-transcribe names from the full 1940 US census.
\end{itemize}

\begin{figure}
    \centering
    \begin{tikzpicture}[x=1cm,y=1cm,>=Stealth]

\pgfmathsetmacro{\mM}{1.0}   
\pgfmathsetmacro{\mO}{0.5}   
\pgfmathsetmacro{\mC}{0.1}   

\pgfmathsetmacro{\xMO}{2.5}  
\pgfmathsetmacro{\xOC}{7.0}  

\pgfmathsetmacro{\aM}{0.0}
\pgfmathsetmacro{\aO}{\aM + (\mM-\mO)*\xMO}
\pgfmathsetmacro{\aC}{\aO + (\mO-\mC)*\xOC}

\pgfmathsetmacro{\yMO}{\aM + \mM*\xMO}
\pgfmathsetmacro{\yOC}{\aO + \mO*\xOC}

\draw[->, very thick] (0,0) -- (0,6.5)
  node[above, align=center] {Total cost of\\research-ready data};
\draw[->, very thick] (0,0) -- (10.5,0) node[right] {Data needed ($N$)};

\draw[thick, dotted] (\xMO,0) -- (\xMO,6.2);
\draw[thick, dotted] (\xOC,0) -- (\xOC,6.2);

\draw[thick, plotblue] (0,\aM) -- (6,{\aM+\mM*6});
\draw[thick, plotgreen] (0,\aO) -- (10,{\aO+\mO*10});
\draw[thick, plotred] (0,\aC) -- (10,{\aC+\mC*10});

\filldraw[plotblack] (\xMO,\yMO) circle (2pt)
  node[above right] {$N_{12}$};
\filldraw[plotblack] (\xOC,\yOC) circle (2pt)
  node[above right] {$N_{23}$};

\node[anchor=west, text=plotblue] at (4.0,5.7) {Manual};
\node[anchor=west, text=plotgreen, align=left] at (7.5,6.5){Off-the-shelf\\(e.g.\ LLM)};
\node[anchor=west, text=plotred, align=left] at (8.6,4.4) {Custom\\solution};

\draw[decorate, decoration={brace, amplitude=7pt, mirror}]
  (0,-0.1) -- (\xMO,-0.1)
  node[midway, below=8pt, align=center] {1. Type it\\yourself};

\draw[decorate, decoration={brace, amplitude=7pt, mirror}]
  (\xMO,-0.1) -- (\xOC,-0.1)
  node[midway, below=8pt, align=center] {2. Off-the-shelf\\solution};

\draw[decorate, decoration={brace, amplitude=7pt, mirror}]
  (\xOC,-0.1) -- (10,-0.1)
  node[midway, below=8pt, align=center] {3. Build your own\\solution};

\end{tikzpicture}
    \caption{Stylized total cost curves for producing research-ready data}
    \label{fig:cost}
    \begin{minipage}{1\linewidth}
        \vspace{5pt}
        \footnotesize
        \textit{Notes:}
        Mode 1 has near-zero fixed costs but high marginal costs; Mode 2 trades higher set-up costs for lower marginal costs; Mode 3 has the highest fixed costs but the lowest marginal costs. The intersection points ($N_{12}$ and $N_{23}$) indicate where the cost-minimizing approach switches.
    \end{minipage}
\end{figure}
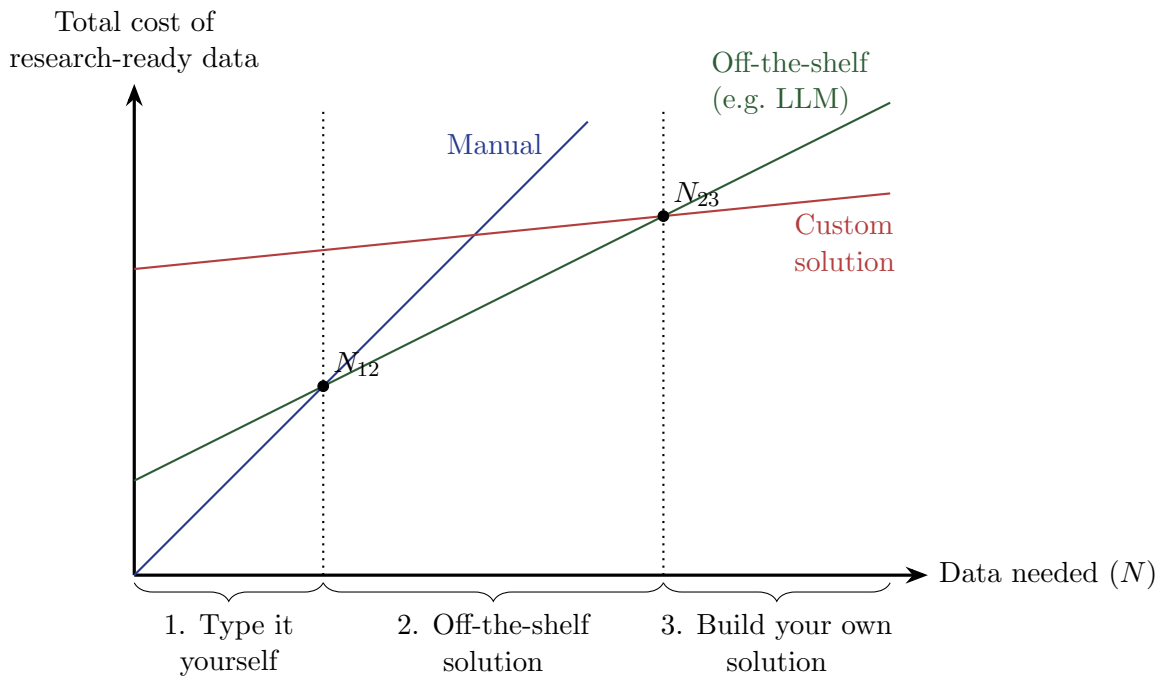

Altogether, we argue that each mode has a useful role---even in the age of highly capable LLMs. The following is a useful heuristic. Write the cost of any data work as
\begin{equation}
C_k(N) = F_k + c_k N,
\end{equation}
where $C_k$ is the total cost of producing \emph{research-ready} data (extracted, cleaned, validated, and documented), $N$ is the amount of data needed, $F_k$ is the fixed cost (learning, scripting, prompt and pipeline design, evaluation set-up), and $c_k$ is the marginal cost (typing or inference time, post-processing, and quality assurance per unit). Only $F_k$ and $c_k$ vary across methods, and both are set by the available technology, researcher skill, infrastructure, and the task's inherent difficulty. Figure~\ref{fig:cost} plots the three modes.

The intersection points in Figure~\ref{fig:cost} ($N_{12}$ and $N_{23}$) are where the cost-minimizing mode switches. These cannot be pinned down universally, but a rough rule of thumb helps: Tasks requiring hours to tens of hours are usually best done in Mode 1. Hundreds to thousands of hours justify Mode 2, and thousands of hours or more may benefit from Mode 3. One further consideration tips toward Mode 3. A custom model released openly, becomes a Mode~2 tool for the rest of the field, so its fixed cost is paid only once, as with the historical BERT model \textit{MacBERTh} \citep{MacBERTh_2021} or the occupational coder \textit{OccCANINE} \citep{dahl2026}, each is built once and reused across many later studies.

Two caveats are important. First, ``research-ready'' is not just extraction. It includes error detection, validation, reproducible documentation, and \textit{debiasing} where needed. This leaves every project with an irreducible manual component: Even a fully automated pipeline needs a hand-labeled gold-standard sample to validate and debias its output, so no project is purely Mode 2 or Mode 3. Second, the cost of automation is hard to predict ex ante. Unanticipated failure modes (noisy scans, domain shifts, ambiguous entity references) can force reruns, further model iterations, or partial redesign. Both raise the effective fixed cost of automation, so it typically becomes worthwhile at a larger scale than it first appears---and it is often worth typing manually for larger projects than one would expect. 

\FloatBarrier
\subsection{New and old approaches: \textit{Bag-of-words} and transformer models}
\label{subsec:bag_of_words}

Bag-of-words models represent a document by the counts, or weighted counts as in \textit{tf-idf}, of the words it contains, discarding word order and context. They are cheap: They do not require a pre-trained model and scale easily to large corpora, which keeps them common in the literature. The cost of discarding word order depends on the task. On the ambiguity scale of Figure~\ref{fig:ambiguity_scale}, tasks at the left end---counting figures, matching standardized strings---lose little, because word identity carries the relevant information. Tasks toward the right lose more, as the discarded context becomes the information of interest. This affects e.g. synonymy, polysemy, stylistic variation, or irony. Most historical text falls toward the right.

This matters for bias, not only accuracy. A simpler model is not a more neutral one. Because bag-of-words representations cannot read a word in light of its neighbors, their errors track the same features that drive systematic bias, so the validation and debiasing of Section~\ref{sec: debiasing} apply to them exactly as to transformers. The motivating example of Section~\ref{sec: debiasing-example} is itself a bag-of-words classifier. The practical question is therefore not whether bag-of-words methods are biased---they and transformers both are---but which starts from a higher level of accuracy. On historical text the evidence favors transformers: \cite{dell2023american} find that fine-tuned BERT models and LLMs outperform bag-of-words approaches on newspaper classification, and \cite{carlson2025unifying} show that for geopolitical risk \citep{caldara2022measuring} a bag-of-words measure can substantially understate the true coefficient. We therefore recommend transformer models where resources allow and debiasing in either case.

\FloatBarrier
\subsection{Digitization}

Large language models (LLMs) and visual large language models (VLMs) have transformed the digitization of historical sources. Until recently, researchers who wanted to extract structured information from text had to rely on optical character recognition (OCR) together with rules for extraction. Such rules, for example from regex commands, worked well for highly structured, repetitive datasets such as matriculation registers \citep{Koschnick2025}, standardized lists of scientists \citep{arora2025rise}, or printed probate records \citep{cummins2019middle}. More complex texts like biographies did not have such tools and were processed by hand \citep{COX2025}. Tables and other visually structured material such as newspapers defied OCR as well, forcing researchers onto more tailored approaches \citep[see, e.g.,][]{dahls2019,shen2021layoutparser,correia2023digitizing}. When text was handwritten, adequate transcription accuracy required large labeled datasets of handwritten names or dates \citep{dahl2023hana,dahl2026dare}.

LLMs, such as the GPT or Llama series, made it possible to process much more complex text. \cite{lagakos2025american} and \cite{ford2025origins} use LLMs to extract detailed information from biographies and life histories, drastically reducing the cost of processing large corpora of semantically complex text. LLMs, however, helped little with visually demanding structures such as tables.

Visual LLMs (VLMs) such as Google's \textit{Gemini} closed that gap. Unlike LLMs, which take text as input, VLMs process images directly, handling the visual structure and the content of a text at once. This can sidestep the OCR and table-recognition step: a scanned image is passed to the VLM together with a prompt describing the layout and the desired output.

VLMs can therefore process \textit{tabular}, \textit{semi-structured}, and \textit{free-flowing} text.\footnote{Researchers might also consider emerging agentic tools such as \textit{Chronos}, which supports source-specific VLM extraction workflows for scanned historical documents and helps researchers adapt these workflows through natural-language interaction \citep{hufe2026chronos}.} Even for free-flowing text they can be better at extracting and aligning elements such as footnotes or marginal comments. On historical corpora they often outperform standard OCR \citep{dasanaike_zeroshot_2026}, and for German patents, 1877--1918, \cite{griesshaber2025multimodalllmshistoricaldataset} provide first evidence that a VLM can achieve a lower error rate than research assistants. VLMs have their own problems, however, notably hallucinated numbers and formatting.

\begin{figure}[ht]
\centering
\begin{tikzpicture}[
  font=\normalsize,
  >=Latex,
  arr/.style={->, line width=1pt},
  box/.style={draw, rounded corners=3pt, align=center, inner sep=5pt},
  txt/.style={align=center},
  lbl/.style={font=\small, align=center}
]

\node[txt] (scan) {Scan};

\node[txt, below left=1.cm and 2cm of scan] (hand) {handwritten};
\node[txt, below right=1cm and 2cm of scan] (machine) {machine\\written};

\node[txt, below=1.1cm of hand] (htr) {HTR};
\node[txt] (ocr) at ([xshift=7.1cm]htr) {OCR};

\node[box, below=4cm of scan] (vlm) {VLM};
\node[box, below=1.1cm of vlm] (llm) {LLM};
\node[txt, below=1.2cm of llm] (out) {Output};

\draw[arr] (scan) to[bend right=25] (hand);
\draw[arr] (scan) to[bend left=25]  (machine);

\draw[arr] (hand) -- (htr);
\draw[arr] (machine) -- (ocr);

\draw[arr] (htr) to[bend right=18] (vlm.west);
\draw[arr] (ocr) to[bend left=18]  (vlm.east);

\draw[arr, dashed] (ocr) to[bend left=22] (llm.east);
\draw[arr, dashed] (htr) to[bend right=22] (llm.west);

\draw[arr] (vlm) to[bend right=35] (out.west);

\draw[arr] (llm) -- (out);

\draw[arr] (htr) to[bend right=55] (out.west);
\draw[arr] (ocr) to[bend left=55]  (out.east);

\end{tikzpicture}
\caption{Pipeline for digitizing scans in economic history}
\label{fig:digitization_worfklow}
\begin{minipage}{1\linewidth}
        \vspace{5pt}
        \footnotesize
        \textit{Notes:}
        The figure illustrates different options researchers can choose when using machine learning for the digitization of scanned sources. To obtain text, researchers can employ either handwritten text recognition (HTR) or optical character recognition (OCR), depending on the source material. The output can then be passed either to a visual large language model (VLM) or to a classic large language model (LLM). These models then extract the information of interest.
    \end{minipage}
\end{figure}
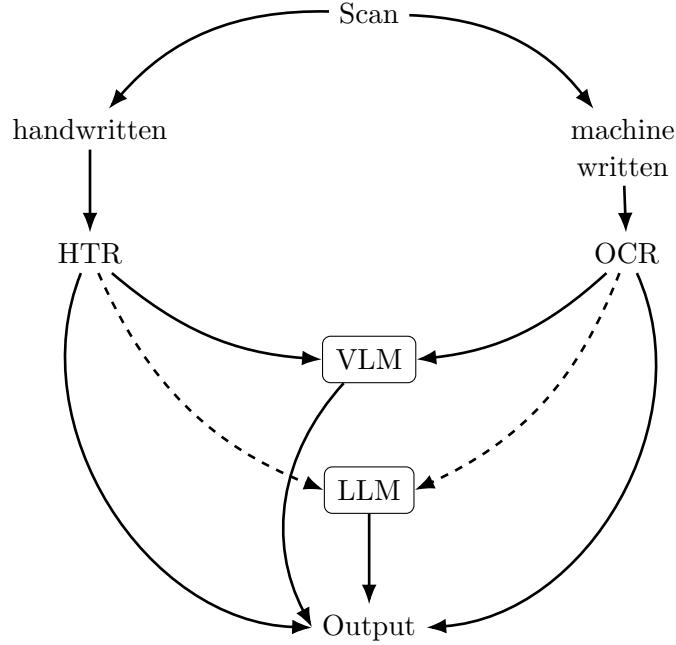

To address this challenge, we recommend combining OCR and VLMs for visually demanding text. This approach directly addresses the problem of hallucinations from VLMs and can achieve significant performance gains over VLM-only approaches \citep{bhaskaran2025automating,greif2025multimodalllmsocrocr}. In a first step, a researcher extracts a precise OCR reading of the text using state-of-the-art tools such as \textit{tesseract} or \textit{aws textract}. Then she passes both the OCR text and the image to a VLM. The VLM is then prompted never to use text snippets or numbers other than those present in the OCR, and to use the visual information from the image to extract tables or other visually structured information.

We illustrate this workflow in Figure~\ref{fig:digitization_worfklow}. Here, an economic historian can first choose either optical character recognition for machine-written text or handwritten text recognition (HTR) for handwritten historical text. Then, in the presence of visually structured text, the text and image are passed into a VLM. Alternatively, researchers can choose to pass the HTR or OCR output into a classical LLM (dashed line) if the text does not include visually structured information. Likewise, in some instances, the HTR or OCR output can be the final output if the researcher is interested in analyzing full text.

Finally, researchers should actively validate the model output against their original sources. While the approach can yield almost perfect accuracy in some settings \citep{backerperal2025llmscrediblytransformcreation,greif2025multimodalllmsocrocr, griesshaber2025multimodalllmshistoricaldataset}, it might still underperform in other settings. Output quality is also highly dependent on the adequacy of the prompting. To improve quality, researchers might also choose to combine multiple VLMs or LLMs in a multi-agent approach, where LLMs can act as judges of the quality of the output of other LLMs \citep{zheng2023judging}.

\subsection{Reproducibility with LLMs}
Lastly, we offer advice on reproducibility when using LLMs. This turns out to be a more complex problem than most researchers would expect.

In contrast to most statistical tools used in economics, which rely on deterministic CPU-based computation with fixed numerical pipelines, large language models operate on stochastic GPU-based architectures. As a result, standard LLM implementations do not guarantee fully deterministic outputs, even when the sampling temperature is set to zero \citep{barrie2024replication}. Although several providers, including \textit{OpenAI}, offer seeding options that ensure a high likelihood of reproducibility, these do not ensure exact and universal reproducibility.\footnote{The OpenAI documentation explicitly states that ``determinism is not guaranteed'' and that users should expect ``almost always the same result'' (\url{https://cookbook.openai.com/examples/reproducible_outputs_with_the_seed_parameter}, accessed 8/15/2025). See also \cite{korinek2023generative}.} Reproducibility is therefore affected by two distinct sources of nondeterminism. First, inference relies on randomized sampling procedures, which can generate different outputs even under identical prompts and hyperparameters. Second, GPU execution itself introduces numerical nondeterminism through parallel floating-point operations, non-associative reductions, and hardware-level scheduling, which can lead to small perturbations that propagate through deep networks.

This problem arises from floating-point arithmetic on GPUs \citep[see, e.g.,][]{yuan2025understandingmitigatingnumericalsources}. Floating-point addition is \emph{not associative}: $(a+b)+c \neq a+(b+c)$ whenever rounding occurs. Because numbers are stored with finite precision, the order in which additions are carried out affects the rounded result. For instance, $(10^{16}+1)+1$ rounds to $10^{16}$, while $10^{16}+(1+1)$ rounds to $10^{16}+2$. On GPUs, large tensor operations are computed in parallel, meaning that the order of elementary additions depends on thread scheduling and hardware-level optimizations. Two runs of the same model may therefore sum the same values in a different order, leading to small numerical differences in intermediate activations. These differences then pass through nonlinear functions (e.g., softmax), where they can be amplified. As a result, models may not produce bit-identical outputs across runs, even with identical inputs and seeds.

Recent approaches have proposed methods for solving this problem, and it seems to be possible to make GPU operations with LLMs deterministic \citep{yuan2025understandingmitigatingnumericalsources,he2025nondeterminism}. Yet, as researchers, we should be aware that these approaches have considerable disadvantages. First, they force researchers to use a temperature of zero, which often produces worse output than slightly higher temperatures. Second, these adjustments come with considerable decreases in computational speed.\footnote{Appendix Section~\ref{app:sec:replicability} provides a user guide to adjusting standard LLM settings to disable various standard features, such as autotuners, that are likely to break exact reproducibility; disabling these features is likely to increase performance losses further. These are considerable downsides. Researchers willingly forego both the best available output from LLMs and large-scale projects that would break their computing budget under exact reproducibility.}

\begin{table}[htbp]
\centering
\begin{threeparttable}
\caption{Reproducibility, Replicability, Robustness, and Generalizability}
\label{tab:repro}
\renewcommand{\arraystretch}{1.5}
\setlength{\tabcolsep}{10pt}
\begin{tabular}{@{}lcc@{}}
\toprule
 & \textbf{Same data} & \textbf{Different data} \\
\midrule
\textbf{Same method}      & \textit{Reproducibility} & \textit{Replicability} \\
\textbf{Different method} & \textit{Robustness}      & \textit{Generalizability} \\
\bottomrule
\end{tabular}
\begin{tablenotes}[flush]
\footnotesize
\item \textit{Notes:} This is the framework introduced by \cite{schloss2018_2x2} to distinguish different forms of replication evidence. See \cite{dreber2025rep} for a discussion of this sort of framework in economics.
\end{tablenotes}
\end{threeparttable}
\end{table}

Yet exact reproducibility is not the ultimate goal of empirical research. It is common to distinguish between reproducibility, replicability, robustness, and generalizability, depending on whether a result is re-derived with the same or different data and the same or different methods (Table~\ref{tab:repro}). The ultimate scientific aim is often generalizability: Results that hold across data and methods. Seen in this light, the small perturbations a GPU introduces do little damage. Insisting on bit-identical output, by contrast, is a standard inherited from deterministic CPU computation that makes little sense in the age of GPU-based models.

We therefore propose an alternative standard, which we call \textit{$\varepsilon$-reproducibility}. Running the same LLM pipeline twice produces almost identical output, with deviations of order $\varepsilon$. In this respect, \textit{$\varepsilon$-reproducibility} should parallel a conscientious research assistant who would label the data almost exactly the same way after being shown it many times, but would still sometimes deviate from their previous output. Indeed, the field of economics and economic history has long worked with RA-extracted data. While this might have led to problems, such as poor coding quality or a lack of historical context, exact reproducibility was never perceived as a shortcoming of RA-labeled or -extracted data.

We expect that the main costs will arise in the editorial process. Exact reproducibility is easy to check, while \textit{$\varepsilon$-reproducibility} might require more care when comparing output. Yet this might appear to be more challenging than it is. We argue that in any common-sense scenario, deviations from running LLMs under \textit{$\varepsilon$-reproducibility} should only lead to minimal changes in downstream tasks, most likely at later decimal places in regression coefficients. If models produce highly varying output after minimal changes to their input, the problem is not \textit{$\varepsilon$-reproducibility} of the LLM but the overall stability of the downstream model. Thus, a common-sense standard might be to require that downstream statistics or coefficients based on LLMs should not deviate by more than, e.g., 1\% of the number presented in the original paper. Editors can also choose to adopt a lower threshold.

Importantly, \textit{$\varepsilon$-reproducibility} does not introduce inexactness into the results or their robustness. Complete exactness of LLM-generated downstream statistics is in any case an illusion, since using other, equally plausible models or slightly different temperatures is likely to lead to much larger variations (which is why it is important to report robustness to such changes). Verification of results by re-running the code remains entirely feasible, and the associated performance gains make reproducibility easier for other research teams. The only true cost may be deeper engagement with the replication material in the editorial process, which can easily be limited by simple threshold rules. We therefore argue that the discipline should be prepared to bear this cost and to abandon \textit{exact reproducibility} in favor of \textit{$\varepsilon$-reproducibility} for LLM-based research.
Abandoning exact reproducibility also means that researchers will be free to focus on more serious replication, robustness, and generalizability tasks.

Finally, we suggest adopting local models, since commercial providers are likely to change their underlying models at some point in the future. Although some providers provide timestamps, they openly acknowledge that this does not guarantee that all parts of the model architecture will remain fixed over time. Additionally, we cannot predict when commercial providers will retire old models. Such changes would undermine both \textit{exact}- and \textit{$\varepsilon$-reproducibility}. Therefore, we follow \cite{ollion2024dangers} and \cite{ludwig2025large} in recommending the use of locally run open-source models, such as Meta's Llama series. Such models should be saved locally and could be included in the replication package.

\section{Conclusion}
\label{sec:conclusion}

This paper has presented an overview of how much ML tools have already transformed the field of economic history. It has then highlighted the danger of bias in downstream models from structural errors in ML predictions. Naively applying ML tools to produce variables that are then used in downstream tasks can lead to severely biased estimates and false research findings. This problem is especially pressing for the field of economic history where the scarcity of historical data often leaves the use of ML predictions as the only viable way to answer many research questions.

However, we argue that there are productive ways of addressing bias from ML predictions in applied work. The most central remedy here is the adoption of a debiasing framework \citep{angelopoulos2023prediction, egami2023using, egami2024using, ludwig2025large, carlson2025unifying}. Here, the researcher uses ML tools to label their data and then draws a small, randomly sampled subset for which historically trained experts produce gold-standard labels. Comparing the predictions against these labels identifies the bias in the ML output, which is then used to correct the downstream estimate. The resulting estimator remains consistent while retaining much of the statistical power from the full set of ML predictions.
We argue that most problems in economic history can be addressed using this debiasing approach. Only cases where ML is used to construct fundamentally new measures invalidate debiasing. Here, careful validation against other proxies and placebos remains the best alternative in such settings.

To help researchers determine which approach fits their problem, we have constructed a simple taxonomy of ML-tasks in economic history. We distinguish between three types. First we define tasks (Type~1) where ML replaces human annotators, and where a gold standard can always be constructed by relying on human experts. Most applications, such as sentiment analysis, record classification, or linkage, fall under this category. For the second task (Type~2), ML is used to address missing data. Here, two possibilities arise. Either researchers can validate a subset of predictions against the source, in which case Type~2 tasks can be represented as Type~1 tasks and sufficiently addressed within the debiasing framework. Alternatively, in cases where researchers are unable to directly verify a subset of their source, Type~2 tasks should be represented as Type~3 tasks. Type~3 tasks are cases where researchers employ ML tools to construct fundamentally new measures of economic phenomena that were previously unavailable. By definition, it is impossible to create a gold standard validation dataset for these types of problems. Therefore, we recommend that researchers adopt a careful validation approach based on the logic of surrogate measures \citep{athey2025surrogate}. We argue that researchers should demonstrate that the new ML predicted measures should strongly correlate with multiple established proxies while remaining unrelated to conceptually similar but economically distinct outcomes.


We have further argued that a key element of the suggested debiasing framework consists in the construction of a gold standard dataset. Here we have argued for the importance of the \textit{critical historian} within the workflow. Debiased estimates can only be as good as the gold standard dataset created by human labelers. Therefore, it is important that the labeling operation is not only conducted carefully but also performed within a broad historical frame that includes source criticism and a careful evaluation of the time domain at hand. In this framework, the use of ML tools is not a substitute but a complement to careful historical work.

The paper then closes by discussing best practices when applying ML tools in economic history. First, we provide guidance on the practical application of the debiasing framework. Next, we consider the choice between off-the-shelf models, fine-tuning, and completely custom-built solutions. Finally, we discuss the problem of exact reproducibility when using LLMs and argue for the adoption of near-exact \textit{$\varepsilon$-reproducibility} instead of exact replicability as a new standard for the profession. We close the paper by discussing recent advances in using ML tools for digitizing data.

Lastly, we note that the future of ML developments remains exciting. We are confident that new methods will further expand the research frontier of economic historians. Recent developments include pre-trained LLMs with a clear historical cutoff \citep{goettlichetal2025,underwood2025,varnum2024large} and agentic models \citep{korinek2025ai_agents}. Yet, we also hope that the potential of ML model-induced bias will remain an important consideration in the use of future innovations in ML. Better tools will not relieve us of the duty to treat them critically.

\clearpage

\bibliographystyle{apacite}
\bibliography{bibliography}

\clearpage
\appendix
\appendixpage
\setcounter{table}{0}
\renewcommand{\thetable}{A\arabic{table}}	
\setcounter{figure}{0}
\renewcommand{\thefigure}{A\arabic{figure}}
\pagenumbering{roman}
\section{Naive Bayes Gender Classifier}
\label{sec:nb_classifier}

We classify gender from first names using a Naive Bayes model trained on
Scandinavian census records from around the same period as the Stockholm data.
The training set combines 1,000 name--gender pairs drawn from the Danish 1880
Copenhagen census \citep{mathiesen2022linklives} and 1,000 pairs from the 
Swedish 1880 IPUMS census extract \citep{IPUMS}, giving 2,000 observations in total.
Both sources record biological sex alongside the full given name; we retain the
first name only, lowercased.

The model treats each first name as a single categorical feature and estimates
the class-conditional probability $P(\text{name} \mid \text{female})$ and
$P(\text{name} \mid \text{male})$ from the training frequencies, with Laplace
smoothing (pseudocount of 1) to handle unseen names.
At prediction time we apply a flat prior $P(\text{female}) = 0.50$, reflecting
uncertainty about the sex composition of the Stockholm taxpayer register.
Names that appear in the Stockholm data but not in the training vocabulary are
assigned a uniform likelihood over all training names.
A taxpayer is classified as female whenever the posterior probability exceeds
0.5.

Table~\ref{tab:confusion} reports classifier performance evaluated on the 6,243
Stockholm observations for which true sex is recorded.
Overall accuracy is 81.0 percent, but errors are strongly asymmetric.
Precision for the female class is 97.6 percent while recall is only 67.2 percent
(F1 = 79.6 percent): the model almost never labels a man as a woman, but
misclassifies 32.8 percent of women as men.

\begin{table}[h]
\caption{Naive Bayes classifier performance, Stockholm 1920 taxpayers
         ($n = 6{,}243$ observations with recorded sex).}
\label{tab:confusion}
\centering
\begin{threeparttable}
\begin{tabular}{lrr}
\toprule
                     & Actual female & Actual male \\
\midrule
Predicted female     & 2,310         & 57          \\
Predicted male       & 1,129         & 2,747       \\
\midrule
Accuracy             & \multicolumn{2}{c}{81.0\%}  \\
Precision (female)   & \multicolumn{2}{c}{97.6\%}  \\
Recall (female)      & \multicolumn{2}{c}{67.2\%}  \\
F1 (female)          & \multicolumn{2}{c}{79.6\%}  \\
\bottomrule
\end{tabular}
\begin{tablenotes}
\small
\item \textit{Note:} Precision, recall, and F1 are computed for the female
class. The training data consist of 1,000 Danish and 1,000 Swedish name--gender
pairs from the respective 1880 censuses.
\end{tablenotes}
\end{threeparttable}
\end{table}

Table~\ref{tab:dsl_results} reports the four wage-gap estimates discussed in
Section~\ref{sec: debiasing-example}.
The na\"{i}ve ML estimate attenuates the gap relative to the oracle by 10.9
log-points.
The DSL estimate recovers the oracle point estimate almost exactly
($-0.814$ vs.\ $-0.822$) while remaining within the oracle confidence interval,
using only a 10 percent gold-standard sample (330 observations).

\begin{table}[h]
\caption{Gender wage gap estimates, Stockholm 1920.
         Outcome: log labor income ($n = 3{,}271$).
         Gold-standard sample: 10\% of observations (330 observations).}
\label{tab:dsl_results}
\centering
\begin{tabular}{lrrl}
\toprule
Estimator            & Estimate & SE    & 95\% CI              \\
\midrule
Na\"{i}ve (ML label) & $-0.713$ & 0.028 & $[-0.767,\ -0.658]$ \\
Labeled-only         & $-0.760$ & 0.076 & $[-0.908,\ -0.612]$ \\
DSL                  & $-0.814$ & 0.048 & $[-0.909,\ -0.719]$ \\
Oracle (true sex)    & $-0.822$ & 0.025 & $[-0.871,\ -0.774]$ \\
\bottomrule
\end{tabular}
\end{table}

\section{Prediction-Powered Inference vs Design-based Supervised Learning: Missing LHS Variable}
\label{sec: ppi-vs-dsl}

Prediction-powered inference (PPI) and design-based supervised learning (DSL) differ slightly in how they obtain unbiased estimates in the presence of biased ML predictions.
To illustrate this difference, consider the example of estimation of the mean discussed in Section~\ref{sec: debiasing-ml-preds}, which we centered around PPI.
We consider the same setting of estimating the mean $\theta$ of $Y$ and show how PPI and DSL differ from each other.

Consider a sample of size $N$.
We observe $S_i = (R_iY_i, X_i, R_i)$ where $Y_i \in \mathbb{R}$ is our quantity of interest, $X_i \in \mathbb{R}^k$ are features an ML model uses to predict $Y_i$, and $R_i \in \{0, 1\}$ is an indicator which equals 1 if we observe $Y_i$, i.e., it denotes our ``gold standard'' subset and we have $n := \sum_{i=1}^NR_i \ll N$.
Further, we have access to an ML model $f: \mathbb{R}^k \rightarrow \mathbb{R}$ which produces predictions $\hat{Y}_i = f(X_i)$.
For notational convenience, assume we order the observations such that $R_i = 1$ for all $i \leq n$ and 0 otherwise.
Since our ``gold standard'' dataset is a random subsample, we have $Y_i \ind R_i$.

\subsection{PPI}

In PPI, we use the estimator

\begin{align*}
    \hat{\theta}_{\text{PPI}} = \frac{1}{N - n} \sum_{i = n + 1}^N f(X_i) - \frac{1}{n} \sum_{i = 1}^n \left( f(X_i) - Y_i \right),
\end{align*}
where the first term, $\frac{1}{N - n} \sum_{i = n + 1}^N f(X_i)$, is the average value of our ML predictions for the sample without ``gold standard'' labels, i.e., where $R_i = 0$.
The second term, $\frac{1}{n} \sum_{i = 1}^n \left( f(X_i) - Y_i \right)$, is the average value of our prediction errors for the sample with ``gold standard'' labels, i.e., where $R_i = 1$.

\subsection{DSL}

In DSL, we first construct pseudo outcomes using our predicted values for both the samples with and without ``gold standard'' labels

\begin{align*}
    \tilde{Y}_i := f(X_i) + \frac{R_i}{\pi}\left( Y_i - f(X_i) \right),
\end{align*}
where $\pi \in (0, 1]$ is the probability we add a ``gold standard'' label to an observation; thus $\pi$ is roughly equal to the share of observations for which $R_i = 1$.
Note how the second term, $\frac{R_i}{\pi}\left( Y_i - f(X_i) \right)$, at first appears problematic, since we do not observe $Y_i$ for all $i = 1, 2, \dots, N$; however, since we multiply by $R_i$, the problematic instances just equal 0.
As such, the second term only affects $\tilde{Y}_i$ when $R_i = 1$, i.e., for our ``gold standard'' labels.
Its denominator then ensures the correction is ``large enough'' to affect the average of $\tilde{Y}_i$.
For example, if $\pi = 0.1$, $\pi$ ensures we ``amplify'' the correction with a factor of 10 for those cases where $R_i = 1$.

Once we have our pseudo outcomes, $\tilde{Y}_i$, DSL uses the estimator

\begin{align*}
    \hat{\theta}_{\text{DSL}} = \frac{1}{N} \sum_{i = 1}^N \tilde{Y}_i,
\end{align*}
which is just the average of the pseudo outcomes for the full sample.

\subsection{Estimator Variance}
\label{sec: ppi-vs-dsl-variance}

The variance of our PPI estimate of $\theta$ is

\begin{align*}
    \mathbb{V}\left[\hat{\theta}_{\text{PPI}}\right] = \frac{\hat{\sigma}^2_{f(X) - Y}}{n} + \frac{\hat{\sigma}^2_{f(X)}}{N-n},
\end{align*}
where $\hat{\sigma}^2_{f(X) - Y}$ is the variance of our residuals on the sample with ``gold standard'' labels (i.e., observations for which $R_i = 1$) and $\hat{\sigma}^2_{f(X)}$ is the variance of our ML predictions on the sample without ``gold standard'' labels (i.e., observations for which $R_i = 0$).
We can thus construct a 95\% confidence interval as

\begin{align*}
    \hat{\theta}_{\text{PPI}} \pm 1.96\sqrt{\mathbb{V}\left[\hat{\theta}_{\text{PPI}}\right]}.
\end{align*}

The variance of our DSL estimate of $\theta$ is
\begin{align*}
    \mathbb{V}\left[\hat{\theta}_{\text{DSL}}\right] = \frac{\hat{\sigma}^2_{\tilde{Y}}}{N},
\end{align*}
where $\hat{\sigma}^2_{\tilde{Y}}$ is the variance of our pseudo outcomes $\tilde{Y}$ on the full dataset.
We can thus construct a 95\% confidence interval as

\begin{align*}
    \hat{\theta}_{\text{DSL}} \pm 1.96\sqrt{\mathbb{V}\left[\hat{\theta}_{\text{DSL}}\right]}.
\end{align*}

Which variance is smaller depends on how many ``gold standard'' labels are available (and is thus closely related to the value of $\pi$ for DSL).
All else equal, DSL will tend to have lower variance relative to PPI when $\pi$ is large.

\subsection{Simulation}

We now show in a simple simulation setting how to use PPI and DSL to estimate the mean of $Y_i$.
Let $N = 1,000$ and $\pi = 0.1$, thus implying $R_i = 1$ for $i \leq n \approx 100$ and $R_i = 0$ otherwise.
Consider the following data generating process

\begin{align*}
    X_i &\sim \mathcal{U}(0, 1) 
    \\
    Y_i &= 2 X_i,
\end{align*}
where $\mathcal{U}(0, 1)$ denotes a uniform distribution with lower bound 0 and upper bound 1.
Clearly, the true mean of $Y_i$ is $\theta = 1$, which we want to  estimate.
We can immediately obtain an unbiased estimate by simply computing the average of $Y_i$ for $i \leq n$, but we want to produce a more precise estimate through the use of an estimator $f(X_i)$ of $Y_i$.
For that purpose, consider the following biased estimator

\begin{align*}
    f(X_i) = 2.1X_i - 0.5X_i^2.
\end{align*}

We now repeatedly draw a sample according to the above, and for each we then compute five quantities
\begin{enumerate}
    \item \textbf{Oracle}: We imagine we observe $Y_i$ for all $N$ samples and compute the average of $Y_i$.
    \item \textbf{``Gold standard''}: We use only observations $i \leq n$ to compute the average of $Y_i$.
    \item \textbf{Naïve ML}: We compute the average of $f(X_i)$ for all $N$ samples.
    \item \textbf{PPI}: We use PPI to debias our ML estimates.
    \item \textbf{DSL}: We use DSL to debias our ML estimates.
\end{enumerate}

\begin{figure}
    \centering
    \includegraphics[width=1\linewidth]{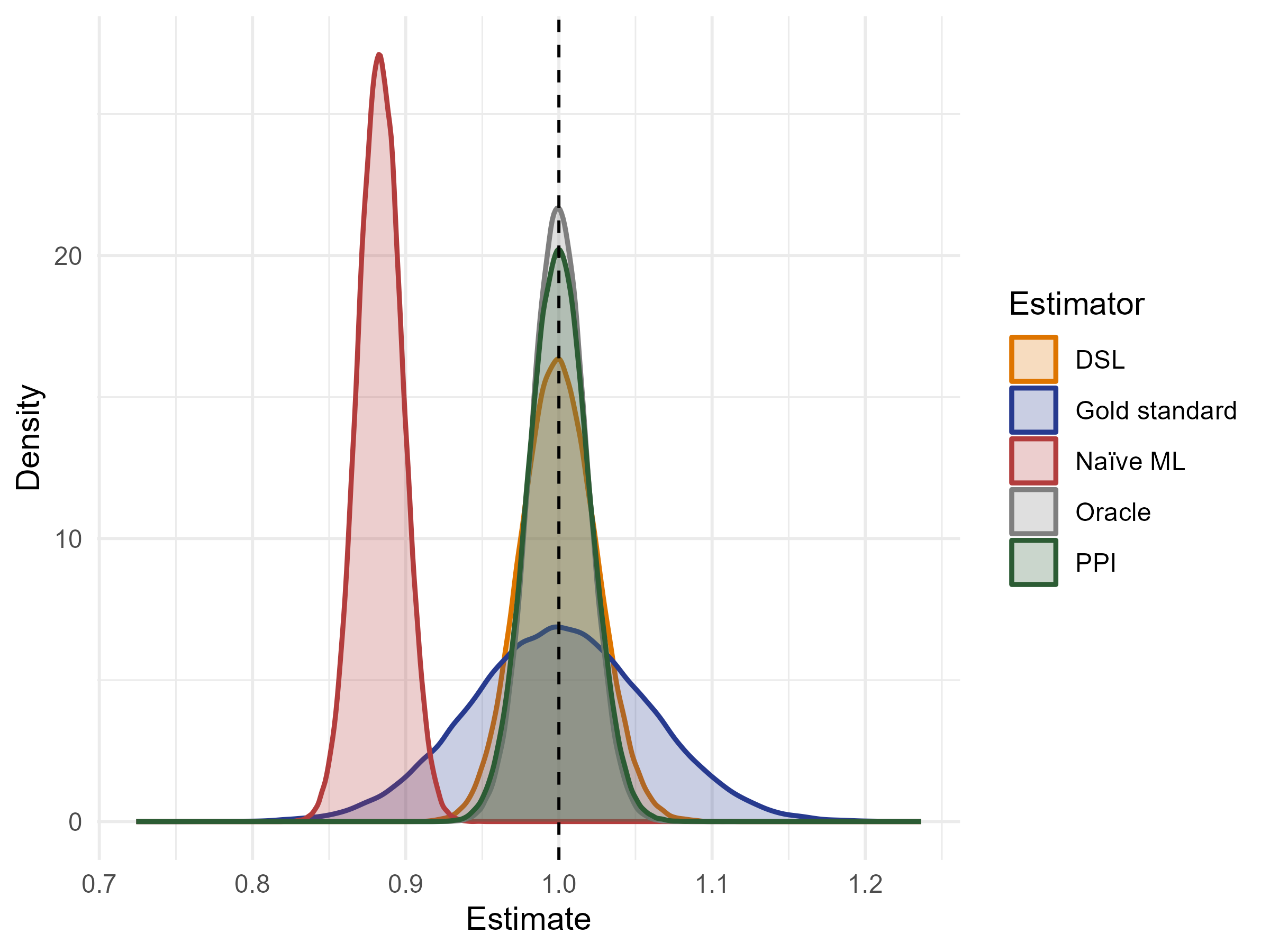}
    \caption{PPI vs DSL for Estimation of a Mean}
    \label{fig: ppi-vs-dsl}
    \begin{minipage}{1\linewidth}
        \vspace{5pt}
        \footnotesize
        \textit{Notes:}
        The figure shows density estimates from five estimators of the mean of $Y_i$ across $100,000$ simulations.
        Oracle refers to an estimator which uses \textit{all} observations of $Y_i$, even when they are not observed.
        ``Gold standard'' refers to an estimator which uses only observations for which we observe $Y_i$.
        Naïve ML refers to an estimator which directly uses ML predictions of $Y_i$ for the full sample, without debiasing.
        PPI refers to the estimator based on prediction-powered inference.
        DSL refers to the estimator based on design-based supervised learning.
        The black, dashed line shows the true mean.
    \end{minipage}
\end{figure}

Figure~\ref{fig: ppi-vs-dsl} shows the results of the above exercise across $100,000$ simulation runs.
Starting with the naïve ML approach, we clearly see it is biased.
Turning to the ``gold standard'' approach, we see it is unbiased, but also that it is relatively imprecise.
The best possible estimator is the oracle estimator, which we see is unbiased and precise; however, since we do not observe $Y_i$ for all $N$ observations, we cannot use this estimator.
However, we see that both PPI and DSL gets us quite close to this theoretical optimum, producing unbiased and precise estimates.

We can use the formulas in Section~\ref{sec: ppi-vs-dsl-variance} for variances of our estimators to construct valid confidence intervals.
Table~\ref{tab: ppi-dsl-ci-coverage} shows the coverage of 95\% confidence intervals for the five above estimators.
As expected, the coverage is close to 95\% for all estimators except the naïve ML estimator of the mean.

\begin{table}[]
    \centering
     \caption{Coverage of 95\% Confidence Intervals}
    \label{tab: ppi-dsl-ci-coverage}
    \begin{tabular}{l r}
        \toprule
        Estimator       & Coverage (\%) \\
        \midrule
        Oracle          & 94.87 \\
Gold standard   & 94.69 \\
Naïve ML       & 0.00 \\
PPI             & 94.87 \\
DSL             & 94.83
\\
        \bottomrule
    \end{tabular}
   \begin{minipage}{1\linewidth}
        \vspace{5pt}
        \footnotesize
        \textit{Notes:}
        The table shows the coverage of 95\% confidence intervals from five estimators of the mean of $Y_i$ across $100,000$ simulations.
        Oracle refers to an estimator which uses \textit{all} observations of $Y_i$, even when they are not observed.
        ``Gold standard'' refers to an estimator which uses only observations for which we observe $Y_i$.
        Naïve ML refers to an estimator which directly uses ML predictions of $Y_i$ for the full sample, without debiasing.
        PPI refers to the estimator based on prediction-powered inference.
        DSL refers to the estimator based on design-based supervised learning.
    \end{minipage}
\end{table}

Taking in sum, these simulations clearly demonstrate the benefits of using debiasing.
We obtain an estimator which is more precise than one which uses only ``gold standard'' labels, while still retaining desirable statistical properties, allowing valid inference.
In the simulations we show, PPI leads to a more precise estimator than DSL, though the two methods are asymptotically equivalent.
Note, however, that this is not the case in general.
If, for example, we increased $\pi$ sufficiently in the above setting, DSL attains lower variance than PPI.

\section{Prediction-Powered Inference vs Design-based Supervised Learning: Missing RHS Variable}
\label{sec: ppi-vs-dsl-rhs}

In some settings, the variable with missing values might be present in the right hand side (RHS) of a regression.
This section details how to use both PPI and DSL in such settings.


Consider a sample of size $N$.
We observe $S_i = (Y_i, R_iZ_i, X_i, R_i)$ where the relationship between $Y_i \in \mathbb{R}$ and $Z_i \in \mathbb{R}$ is our quantity of interest, $X_i \in \mathbb{R}^k$ are features an ML model uses to predict $Z_i$, and $R_i \in \{0, 1\}$ is an indicator which equals 1 if we observe $Z_i$, i.e., it denotes our ``gold standard'' subset and we have $n := \sum_{i=1}^NR_i \ll N$.
Further, we have access to an ML model $f: \mathbb{R}^k \rightarrow \mathbb{R}$ which produces predictions $\hat{Z}_i = f(X_i)$.
For notational convenience, assume we order the observations such that $R_i = 1$ for all $i \leq n$ and 0 otherwise.
Since our ``gold standard'' dataset is a random subsample, we have $(Y_i, Z_i) \ind R_i$.
We assume we are interested in the linear relation, i.e., in estimating $\beta$ in the below equation

\begin{align*}
    Y_i = \alpha + \beta Z_i + \varepsilon_i.
\end{align*}

\subsection{PPI}

Let

\begin{align*}
    \mathbf{Z}_{\text{GS}} = \left(
        \begin{array}{cc}
            1 & Z_{1}       \\
            1 & Z_{2}       \\
            \vdots & \vdots \\
            1 & Z_{n}
        \end{array}
    \right),
    \quad
    \hat{\mathbf{Z}}_{\text{GS}} = \left(
        \begin{array}{cc}
            1 & \hat{Z}_{1} \\
            1 & \hat{Z}_{2} \\
            \vdots & \vdots \\
            1 & \hat{Z}_{n}
        \end{array}
    \right),
    \quad
    \hat{\mathbf{Z}}_{\text{ML}} = \left(
        \begin{array}{cc}
            1 & \hat{Z}_{n + 1} \\
            1 & \hat{Z}_{n + 2} \\
            \vdots & \vdots     \\
            1 & \hat{Z}_{N}
        \end{array}
    \right),
\end{align*}
where $\mathbf{Z}_{\text{GS}}$ is the matrix consisting of a column of ones and the $n$ ``gold standard'' values of $Z_i$ we observe, $\hat{\mathbf{Z}}_{\text{GS}}$ is the matrix consisting of a column of ones and the ML predictions of $Z_i$ for our ``gold standard'' $n$ observations, and $\hat{\mathbf{Z}}_{\text{ML}}$ is the matrix consisting of a column of ones and the ML predictions of $Z_i$ for the remaining $N - n$ observations.

Let $\mathbf{Y}_{\text{GS}} = (Y_1, Y_2, \dots Y_n)^T$ and $\mathbf{Y}_{\text{ML}} = (Y_{n +1}, Y_{n + 2}, \dots Y_N)^T$---the vector consisting of $Y_i$ for the sample where we do and do not observe the values of $Z_i$, respectively.
Using just the ``gold standard'' sample, we could estimate $(\alpha, \beta)^T$ as

\begin{align*}
    \left(
        \begin{array}{c}
            \hat{\alpha}_{\text{GS}}       \\
            \hat{\beta}_{\text{GS}}
        \end{array}
    \right) 
    = \left( \mathbf{Z}_{\text{GS}}^T \mathbf{Z}_{\text{GS}} \right)^{-1} \mathbf{Z}_{\text{GS}}^T \mathbf{Y}_{\text{GS}},
\end{align*}
but we want to use the larger sample combined with ML predictions of $Z_i$ for efficiency.
Note that in the case without any missing data, the optimal estimator is $\left( \mathbf{Z}^T \mathbf{Z} \right)^{-1} \mathbf{Z}^T \mathbf{Y}$.\footnote{Here, $\mathbf{Z}$ is defined the same way as $\mathbf{Z}_{\text{GS}}$, but using all $N$ observations. That is, the two coincide in the case where $n = N$, corresponding to knowing the true label for all observations, which is precisely the oracle case.}

Naïvely using ML predictions, we obtain the (potentially biased) estimator
\begin{align*}
    \left(
        \begin{array}{c}
            \hat{\alpha}_{\text{ML}}       \\
            \hat{\beta}_{\text{ML}}
        \end{array}
    \right) 
    = \left(\underbrace{ \mathbf{Z}_{\text{ML}}^T \mathbf{Z}_{\text{ML}}}_{\mathbf{(1)}}\right)^{-1} \underbrace{\mathbf{Z}_{\text{ML}}^T \mathbf{Y}_{\text{ML}}}_{(2)},
\end{align*}
essentially consisting of two (potentially biased) estimators, namely (1) and (2).
PPI debiases (1) and (2) using true values of $Z_i$ and corresponding ML predictions.
The final estimator is then
\begin{align*}
    \hat{\mathbf{G}} &= \hat{\mathbf{Z}}_{\text{ML}}^T \hat{\mathbf{Z}}_{\text{ML}} / (N - n) - \left(\hat{\mathbf{Z}}_{\text{GS}}^T - \mathbf{Z}_{\text{GS}}^T \mathbf{Z}_{\text{GS}}\hat{\mathbf{Z}}_{\text{GS}}\right) / n
    \\
    \hat{\mathbf{g}} &= \hat{\mathbf{Z}}_{\text{ML}}^T \mathbf{Y}_{\text{ML}} / (N - n) + \left(\hat{\mathbf{Z}}_{\text{GS}}^T \mathbf{Y}_{\text{GS}} - \mathbf{Z}_{\text{GS}}^T \mathbf{Y}_{\text{GS}}\right) / n
    \\
    \hat{\boldsymbol{\theta}}_{\text{PPI}} = \left(
        \begin{array}{c}
            \hat{\alpha}_{\text{PPI}}       \\
            \hat{\beta}_{\text{PPI}}
        \end{array}
    \right) &= \hat{\mathbf{G}}^{-1} \hat{\mathbf{g}}.
\end{align*}

Compared to the simpler setting of missing values for $Y_i$ discussed in Appendix~\ref{sec: ppi-vs-dsl}, this at first looks slightly more complicated.
However, the idea is exactly the same:
We compute an ``average'' of our ML predictions for the sample with missing values of $Z_i$ as well as an ``average'' of the bias of those for the sample where we observe $Z_i$.
The main difference is that we do it twice:
Once in constructing $\hat{\mathbf{G}}$ and once in constructing $\hat{\mathbf{g}}$.
Once we have those two unbiased estimates, we combine them.

\subsection{DSL}

We start by constructing pseudo outcomes of $Z_i$, $Z_iY_i$, and $Z_i^2$ using our predicted values for both the samples with and without ``gold standard'' labels

\begin{align*}
    \tilde{Z}_i &:= f(X_i) + \frac{R_i}{\pi}\left( Z_i - f(X_i) \right) & \text{Predicted values for } Z_i
    \\
    \widetilde{Z_i^2} &:= f(X_i)^2 + \frac{R_i}{\pi}\left( Z_i^2 - f(X_i)^2 \right) & \text{Predicted values for } Z_i^2
    \\
    \widetilde{Z_iY_i} &:= f(X_i) Y_i + \frac{R_i}{\pi}\left( Z_i Y_i - f(X_i) Y_i \right) & \text{Predicted values for } Z_iY_i,
\end{align*}
where $\pi \in (0, 1]$ is the probability we add a ``gold standard'' $Z_i$ to an observation; thus $\pi$ is roughly equal to the share of observations for which $R_i = 1$.

Once we have our pseudo outcomes, the DSL estimator of the slope is

\begin{align*}
    \hat{\beta}_{\text{DSL}} = \frac{\frac{1}{N}\sum_{i = 1}^N \widetilde{Z_iY_i} - (\frac{1}{N}\sum_{i = 1}^N \tilde{Z}_i)(\frac{1}{N}\sum_{i = 1}^N Y_i)}{\frac{1}{N}\sum_{i = 1}^N \widetilde{Z_i^2} - (\frac{1}{N}\sum_{i = 1}^N \tilde{Z}_i)^2},
\end{align*}
which is equivalent to the standard OLS estimator of the slope, but where we now use the pseudo outcomes in place of $Z_i$, $Z_i^2$, and $Z_iY_i$.

\subsection{Estimator Variance}
\label{sec: ppi-vs-dsl-rhs-variance}

\paragraph{PPI}
Let 
$\hat{\mathbf{e}}_{\text{ML}} = \mathbf{Y}_{\text{ML}} - \hat{\mathbf{Z}}_{\text{ML}}\hat{\boldsymbol{\theta}}_{\text{PPI}}$, 
$\hat{\mathbf{e}}_{\text{GS}} = \mathbf{Y}_{\text{GS}} - \hat{\mathbf{Z}}_{\text{GS}}\hat{\boldsymbol{\theta}}_{\text{PPI}}$, and 
$\mathbf{e}_{\text{GS}} = \mathbf{Y}_{\text{GS}} - \mathbf{Z}_{\text{GS}}\hat{\boldsymbol{\theta}}_{\text{PPI}}$.
Let $\circ$ denote the Hadamard product (element-wise multiplication).
Finally, let $\hat{\mathbb{V}}$ denote the (empirical) covariance matrix of a matrix.
The covariance matrix of $\hat{\boldsymbol{\theta}}_{\text{PPI}}$ is then
\begin{align*}
    \hat{\mathbb{V}}[\hat{\boldsymbol{\theta}}_{\text{PPI}}] &= \hat{\mathbf{G}}^{-1} \hat{\mathbf{V}} \hat{\mathbf{G}}^{-1},
    \\
    \hat{\mathbf{V}} &=
    \frac{\hat{\mathbb{V}}\left[\mathbf{Z}_{\text{GS}} \circ \mathbf{e}_{\text{GS}} - \hat{\mathbf{Z}}_{\text{GS}} \circ \hat{\mathbf{e}}_{\text{GS}}\right]}{n} + \frac{\hat{\mathbb{V}}\left[\hat{\mathbf{Z}}_{\text{ML}} \circ \hat{\mathbf{e}}_{\text{ML}}\right]}{N - n}.
\end{align*}


\paragraph{DSL}

We start by estimating the averages of $Y_i$, $Z_i$, $Z_iY_i$, and $Z_i^2$, where we use pseudo outcomes for the last three because $Z_i$ is not observed for all observations.

\begin{align*}
    \hat{\mu}_{Y} &= \frac{1}{N}\sum_{i = 1}^N Y_i, 
    & 
    \hat{\mu}_{Z} &= \frac{1}{N}\sum_{i = 1}^N \tilde{Z}_i,
    \\
    \hat{\mu}_{Z^2} &= \frac{1}{N}\sum_{i = 1}^N \widetilde{Z_i^2},
    & 
    \hat{\mu}_{ZY} &= \frac{1}{N}\sum_{i = 1}^N \widetilde{Z_iY_i}.
\end{align*}

We can thus write $\hat{\beta}_{\text{DSL}}$ as
\begin{align*}
    \hat{\beta}_{\text{DSL}} &= \frac{\overbrace{\hat{\mu}_{ZY} - \hat{\mu}_{Y} \hat{\mu}_{Z}}^{\hat{A}}}{\underbrace{\hat{\mu}_{Z^2} - \hat{\mu}_{Z}^2}_{\hat{B}}},
\end{align*}
with respective estimated influence functions
\begin{align*}
    \mathbb{IF}(\hat{A}) &= \widetilde{Z_iY_i} - \hat{\mu}_{ZY} - \hat{\mu}_{Y} (\tilde{Z}_i - \hat{\mu}_{Z}) - \hat{\mu}_{Z} (Y_i - \hat{\mu}_{Y}),
    \\
    \mathbb{IF}(\hat{B}) &= \widetilde{Z_i^2} - \hat{\mu}_{Z^2} - 2 \hat{\mu}_{Z} (\tilde{Z}_i - \hat{\mu}_{Z}).
\end{align*}
As such, the estimated influence function of $\hat{\beta}_{\text{DSL}}$ is\footnote{By ``estimated influence function'', we refer the plugging in the values of the various quantities in our influence function, such as the estimated means.}
\begin{align*}
    \mathbb{IF}(\hat{\beta}_{\text{DSL}}) &= \mathbb{IF}\left(\frac{\hat{A}}{\hat{B}}\right) = \frac{\mathbb{IF}(\hat{A}) - \hat{\beta}_{\text{DSL}} \, \mathbb{IF}(\hat{B})}{\hat{B}}
    \\
    &= \frac{\overbrace{\left(\widetilde{Z_iY_i} - \hat{\mu}_{ZY} - \hat{\mu}_{Y} (\tilde{Z}_i - \hat{\mu}_{Z}) - \hat{\mu}_{Z} (Y_i - \hat{\mu}_{Y})\right)}^{\mathbb{IF}(\hat{A})} - \hat{\beta}_{\text{DSL}} \overbrace{\left( \widetilde{Z_i^2} - \hat{\mu}_{Z^2} - 2 \hat{\mu}_{Z} (\tilde{Z}_i - \hat{\mu}_{Z}) \right)}^{\mathbb{IF}(\hat{B})}}{\underbrace{\hat{\mu}_{Z^2} - \hat{\mu}_{Z}^2}_{\hat{B}}}
\end{align*}

With the estimated influence function at hand, we know the asymptotic variance of $\hat{\beta}_{\text{DSL}}$ can be estimated by the variance of $\mathbb{IF}(\hat{\beta}_{\text{DSL}})$ divided by the sample size
\begin{align*}
    \hat{\mathbb{V}}[\hat{\beta}_{\text{DSL}}] &= \frac{1}{N} \hat{\mathbb{V}}\left[ \mathbb{IF}(\hat{\beta}_{\text{DSL}}) \right].
\end{align*}


\subsection{Simulation}

We now show in a simple simulation setting how to use PPI and DSL to estimate the slope of the relationship between $Y_i$ and $Z_i$.
Let $N = 1,000$ and $\pi = 0.1$, thus implying $R_i = 1$ for $i \leq n \approx 100$ and $R_i = 0$ otherwise.
Consider the following data generating process

\begin{align*}
    X_i &\sim \mathcal{U}(0, 1) 
    \\
    \varepsilon_i &\sim \mathcal{N}(0, 1) 
    \\
    Z_i &= 2 X_i
    \\
    Y_i &= 1 + 1.5Z_i + \varepsilon_i,
\end{align*}
where $\mathcal{U}(0, 1)$ denotes a uniform distribution with lower bound 0 and upper bound 1 and $\mathcal{N}(0, 1)$ a standard normal distribution.
We can immediately obtain an unbiased estimate by simply regressing $Y_i$ on $Z_i$ for $i \leq n$, but we want to produce a more precise estimate through the use of an estimator $f(X_i)$ of $Z_i$.
For that purpose, consider the following biased estimator

\begin{align*}
    f(X_i) = 2.1X_i - 0.5X_i^2.
\end{align*}

We now repeatedly draw a sample according to the above, and for each we then compute five quantities
\begin{enumerate}
    \item \textbf{Oracle}: We imagine we observe $Z_i$ for all $N$ samples and regress $Y_i$ on $Z_i$ for all $N$ samples.
    \item \textbf{``Gold standard''}: We use only observations $i \leq n$ to perform the above regression.
    \item \textbf{Naïve ML}: We regress $Y_i$ on $f(X_i)$ for all $N$ samples.
    \item \textbf{PPI}: We use PPI to debias our ML estimates.
    \item \textbf{DSL}: We use DSL to debias our ML estimates.
\end{enumerate}

\begin{figure}
    \centering
    \includegraphics[width=1\linewidth]{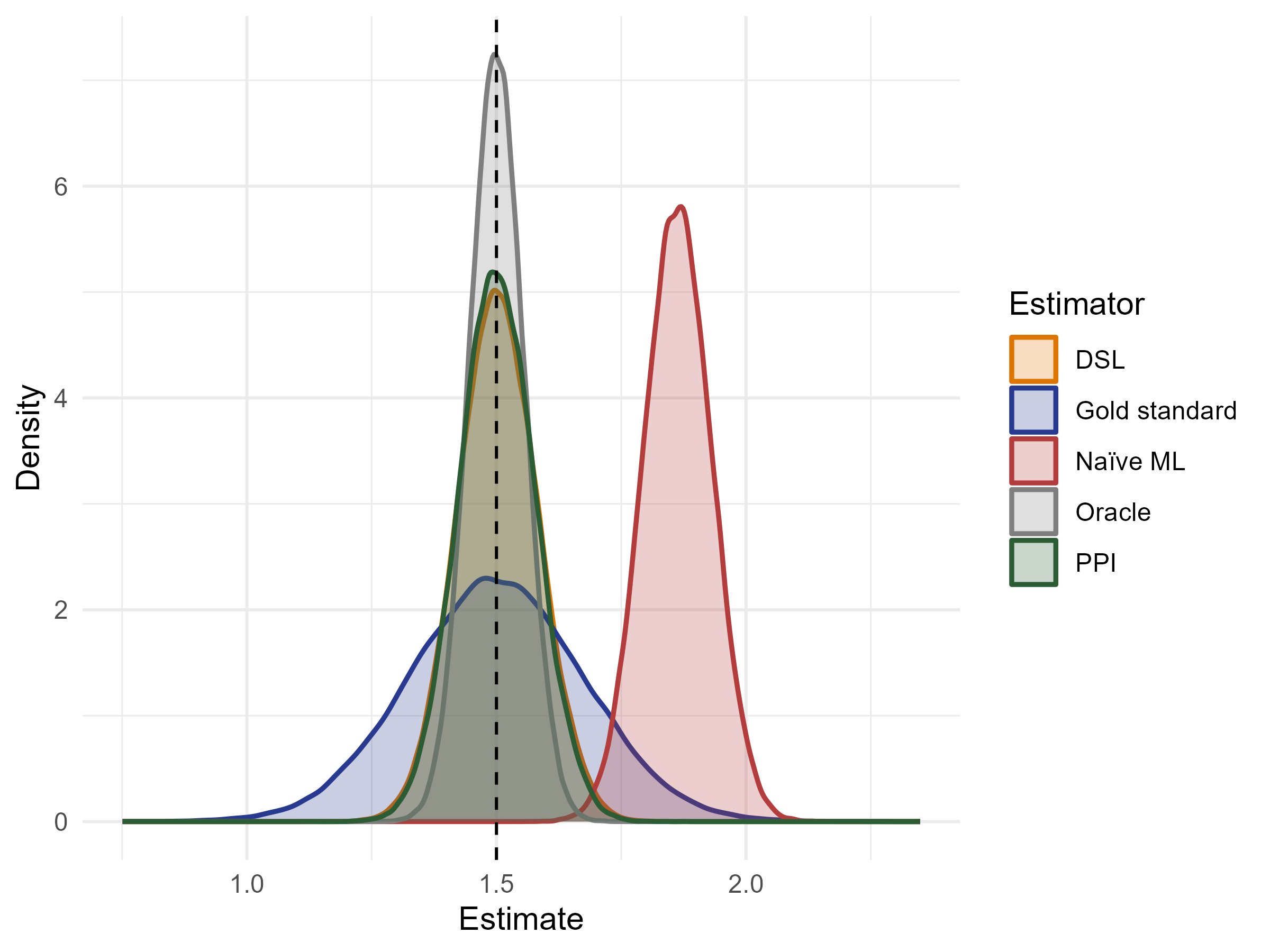}
    \caption{PPI vs DSL for Regression With Missing RHS Variable}
    \label{fig: ppi-vs-dsl-rhs}
    \begin{minipage}{1\linewidth}
        \vspace{5pt}
        \footnotesize
        \textit{Notes:}
        The figure shows density estimates from five estimators of the slope of $Y_i$ with respect to $Z_i$ across $100,000$ simulations.
        Oracle refers to an estimator which uses \textit{all} observations of $Z_i$, even when they are not observed.
        ``Gold standard'' refers to an estimator which uses only observations for which we observe $Z_i$.
        Naïve ML refers to an estimator which directly uses ML predictions of $Z_i$ for the full sample, without debiasing.
        PPI refers to the estimator based on prediction-powered inference.
        DSL refers to the estimator based on design-based supervised learning.
        The black, dashed line shows the true slope.
    \end{minipage}
\end{figure}

Figure~\ref{fig: ppi-vs-dsl-rhs} shows the results of the above exercise across $100,000$ simulation runs.
Starting with the naïve ML approach, we clearly see it is biased.
Turning to the ``gold standard'' approach, we see it is unbiased, but also that it is relatively imprecise.
The best possible estimator is the oracle estimator, which we see is unbiased and precise; however, since we do not observe $Z_i$ for all $N$ observations, we cannot use this estimator.
However, we see that both PPI and DSL gets us quite close to this theoretical optimum, producing unbiased and precise estimates.

We can use the formulas in Section~\ref{sec: ppi-vs-dsl-rhs-variance} for variances of our estimators to construct valid confidence intervals.
Table~\ref{tab: ppi-dsl-rhs-ci-coverage} shows the coverage of 95\% confidence intervals for the five above estimators.
As expected, the coverage matches 95\% for all estimators except the naïve ML estimator of the slope.\footnote{The slight deviations for PPI and DSL occur in part by chance, but also due to a sample size of just 1,000, suggesting only around 100 labeled observations; increasing the sample size leads to results closely matching the asymptotic guarantee of correct coverage.}
Taking in sum, these simulations clearly demonstrate the benefits of using debiasing.
We obtain an estimator which is more precise than one which uses only ``gold standard'' labels, while still retaining desirable statistical properties, allowing valid inference.
In the simulations we show, PPI and DSL have approximately similar variances.

\begin{table}[]
    \centering
     \caption{Coverage of 95\% Confidence Intervals}
    \label{tab: ppi-dsl-rhs-ci-coverage}
    \begin{tabular}{l r}
        \toprule
        Estimator       & Coverage (\%) \\
        \midrule
        Oracle          & 95.03 \\
Gold standard   & 94.82 \\
Naïve ML       & 0.05 \\
PPI             & 94.63 \\
DSL             & 94.39
\\
        \bottomrule
    \end{tabular}
   \begin{minipage}{1\linewidth}
        \vspace{5pt}
        \footnotesize
        \textit{Notes:}
        The table shows the coverage of 95\% confidence intervals from five estimators of the slope of $Y_i$ with respect to $Z_i$ across $100,000$ simulations.
        Oracle refers to an estimator which uses \textit{all} observations of $Z_i$, even when they are not observed.
        ``Gold standard'' refers to an estimator which uses only observations for which we observe $Z_i$.
        Naïve ML refers to an estimator which directly uses ML predictions of $Z_i$ for the full sample, without debiasing.
        PPI refers to the estimator based on prediction-powered inference.
        DSL refers to the estimator based on design-based supervised learning.
    \end{minipage}
\end{table}

\FloatBarrier
\clearpage

\section{Close-reading}
\label{app:sec:close_reading}

When analyzing contradictions between human and LLM labels, it is not ex-ante clear which coding is correct. Therefore, this section takes a small sample of passages where human and LLM labels disagree and proceeds with a brief source criticism and close reading of the source to gauge the meaning of the text in deeper context.  This should serve both as illustration and evidence of how even recent top-models can fail to properly interpret historical sources.

This exercise goes beyond the data shared with the LLM --- as such it is not making the point that human annotators are more capable than an LLM, it simply underscores the point that feeding out-of-context material into an LLM for labeling misses out on much of the historical context and is likely to lead to wrong and/or biased outcomes.  Furthermore, note that often much of the additional context is not accessible to an LLM.  Given that we draw on publicly available books from Early English Books Online (EEBO) material, the original content of the book could have, in principle, been accessed by the LLM. Often, historical material is not digitized and not accessible to LLMs. Moreover, some of the cited secondary literature, such as \cite{schneider2007forde}, are further protected by copyright and paywalls and thus also not accessible to an LLM. All of this together, shows that a full exercise of source criticism and historical deep-reading techniques go beyond the LLMs capabilities in historical research and can lead to much reliable analyses and data codings.

The following discusses two examples of LLM and human disagreement. The first case is a letter extract by \textit{Thomas Forde} published in 1660. The second one is a an extract from Matthew Hopkins vindication, \textit{The discovery of vvitches} published in 1657.

\noindent \textbf{Thomas Forde, 1660}:

We start with the following paragraph published by Thomas Forde in 1660:

\begin{quote}
    By this time, I doubt not, but they who most endeavoured his Majesties death, have seen cause enough to wish him alive again, and are ready to engrave that Motto upon his Statue (which they threw down with contempt) which was set upon the Statue of the Roman Brutus, Utinam viveres. It is yet some comfort that we can mingle sighs, and assist one another with mutual counsels and courtesies, which shall ne ver be wanting from 
\end{quote}

The \texttt{gpt-5.5} LLM assigns the emotions of \textit{anger}, and \textit{sadness} to the text but detects no notion of \textit{irony}. The authors' annotation instead highlights \textit{sadness}, and \textit{irony}. As argued before, it is not ex ante clear, which sentiment labels are correct. We therefore, inspect the source in its context and then interpret its content closely against the historical context. 

The paragraph itself comes from the following book, as listed in the EEBO \textit{Virtus rediviva a panegyrick on our late King Charles the I. \&c. of ever blessed memory. Attended, with severall other pieces from the same pen. Viz. [brace] I. A theatre of wits: being a collection of apothegms. II. Fœnestra in pectore: or a century of familiar letters. III. Loves labyrinth: a tragi-comedy. IV. Fragmenta poetica: or poeticall diversions. Concluding, with a panegyrick on his sacred Majesties most happy return. / By T.F.} (Wing F1548-F1550).

As indicated in the title, the primary book \textit{Virtus rediviva a panegyrick on our late King Charles the I. \&c. of ever blessed memory} is printed together with several other volumes. Its main part, the \textit{Virtus Rediviva}, was written on the anniversaries of Charles I's execution in 1657 and 1658 but legally published following the restoration. It thus is part of the clandestine Royal mourning literature \citep{Lacey_2003,Potter1989}.


The paragraph itself stems from the \textit{Fœnestra in pectore}, a collection of Forde's letters that is attached to the \textit{Virtus rediviva}. The letters are based on Forde's epistolary correspondence between 1642 and the late 1650s but modified for publication in 1660 \citep{schneider2007forde}. The paragraph stems from a letter addressed to Mr. \textit{T. C.} from the late 1650s that reflects back on the execution of Charles I. 

Thomas Forde, born in 1624, was apprenticed to Samual Man at the Stationer's Company and seemed to have been engaged in the book trade by the 1660s \citep{schneider2007forde}. During the civil war, he seems to have mainly retired from his occupation and lived as a private scholar (ibid.). His sympathies were clearly aligned with the Royalist side (ibid.). 

Given this information, we revisit the text itself. The letter originates from the late 1650s, the last years of the interregnum and Commonwealth. The paragraph starts with the claim that many ``who most endeavoured his Majesties death, have seen cause enough to wish him alive again'' --- the letter effectively ascribes a shift of conscience to Forde's ideological opponents in the Parliamentary camp. Forde's own Royalist position becomes eminently clear here. The letter then continues by stating that these former Parliamentarians would now be ready to engrave ``Utinam viveres'' on Charles's statue. It closes with a note how Forde and his Royalist allies can find comfort in morning.

They key piece for interpretation is the \textit{Utinam viveres} description. It comes from Suetonius’s account of the final period of Julius Caesar’s rule.\footnote{Divus Iulius 80.3}:

\begin{quote}
    ``Some people wrote on the satute of Lucias Brutus: Ìf only you were living! and on that of Caesar himself:

    \begin{quote}
        Brutus was made first consul, since he threw out the kings,
        
        He, since he's thrown out the consuls, eventually gets to be king''
    \end{quote}
    (\citeauthor{suetonius2000}, Divius Iulius, 80.3, Catharine Edwards translation)
\end{quote}

It thus captures a classic Roman Republican and distinctly anti-monarchical sentiment. Yet, Forde takes up this anti-monarchical reference and turns it into support for monarchy. Implicitly, the author casts the Commonwealth and Cromwell as tyrants and compares the king to the virtues of the Republic. The phrase is turned powerfully on its head --- reflecting the \textit{world turned upside down} sentiment of the time \citep{hill1972world,healey2024blazing} --- and a fine use of irony that questions the legitimacy of the Commonwealth regime. 

Hence, we can conclude that the two most evocative themes of the letter are both a mourning for Parliament and the direct use of irony that casts doubt on the legitimacy of the Commonwealth. While the LLM is able to identify the emotions of \textit{sadness}, it misses the period-specific, but powerful use of irony. It turns out that by inspecting the source context and direct references in the letter, a critical historian's close reading exercise can yield a better and more time-period appropriate interpretation of the text as an LLM applied to the batch-processing of sentiment in historical text.

\noindent  \textbf{Matthew Hopkins, 1647.}

Next, we continue with another example on top of our random draw. It turns out to come from one of the more infamous figure of the period.

\begin{quote}
    (...) peached one another thereabouts that joyned together in the like damnable practise, that in our Hundred in Essex, 29. were con demned at once, 4. brought 25. Miles to be hanged, where this Discoverer lives, for sending the Devill like a Beare to kill him in his garden, so by seeing diverse of the 〈gap〉 Papps, and trying wayes with hundreds of them, he gained this experience, and for ought he knowes any man else may find them as well as he and his company, if they had the same skill and experience. Quer. 5. Many poore People are condemned for having a Pap, or Teat about them, whereas many People (especially antient People) are, and have been a long time troubled with naturall wretts on severall parts of their bodies, and other naturall excressencies, as Hemerodes, Piles, Childbearing, \&c. and these shall be judged only by one man a lone, and a woman, and so accused or acquitted. 
\end{quote}

The \texttt{gpt-5.5} LLM assigns the emotion of \textit{sadness} to the text but detects no notion of \textit{anger}. The authors' annotation instead highlights \textit{anger} and \textit{fear}.

The passage comes from Matthew Hopkin's (1620-1647) \textit{The discovery of vvitches: in answer to severall queries, lately delivered to the judges of the assize for the county of Norfolk. / And now published by Matthevv Hopkins, witch-finder. For the benefit of the whole kingdome.} (Wing H2751), a short pamphlet published in 1647. 

Matthew Hopkins, is one of the (infamously) better known figures of the seventeenth century. 
Starting as a relative unknown from Essex, Hopkins became a leading figure in the East Anglian witch-hunt of 1644--47. Together with John Stearne, he investigated suspected witches using sleep deprivation, searches for alleged witches' marks, and tests involving supposed familiar spirits. Their campaign contributed to the execution of more than one hundred people, mostly women, and already drew contemporary criticism for its coercive methods \citep{sharpe2004hopkins}.

Published in 1647, \textit{The Discovery of Witches} is a short polemical pamphlet organized as a series of objections and replies. It is is vindicatory in nature, directed at the growing criticism of the coercive practices used during the East Anglian prosecutions \citep{sharpe2004hopkins}. It is structured around numbered queries that object to his methods and his answers to the queries. The paragraph from above mixes the answer to Query 4 with the beginning of Query 5. The paragraph thus constitutes a text that is difficult in structure and incomplete in context. Yet, given the difficulty of obtaining proper text structure and paragraph breaks from historical text, this appears as a common type of example of historical text used for LLM-batch processing.\footnote{The approach used for this LLM-labeling exercise was to simply split paragraphs by line breaks.}

The following recounts the full context of the paragraph at hand. 
Query 3 asks where Hopkins acquired his supposed ability to identify witches. He answers that he gained it from experience. Query 4 then asks where he gained this experience. Hopkins responds by stating that he experienced the case of a witch in his village. She was discovered by her marks ,and kept from sleep for 3 nights, after which she called out her supposed familiars and named various other women in league with her. This is where the paragraph starts as these other women denounced (``peached'') each other.
The paragraph then continues by describing how they were condemned and hanged and recounts how thus, Hopkins, gained his experience. This then followed by Query 5 that questions whether marks (``Pap, or Teat'') can serve to identify a witch when so many people have similar marks on them.

Thus, much of the text, as expected from a vindication, is rather neutral and analytical in tone. 
Yet, the very beginning of the paragraph reveals Hopkins emotional attitude. It contains two distinct emotional positions. Hopkins's answer employs the morally condemnatory phrase ``the like damnable practise'' and recalls the alleged attempt to kill him by sending the Devil in the shape of a bear. These formulations communicate hostility towards the accused and fear of a purported demonic threat. They provide a clear basis for the labels \textit{anger} and \textit{fear}. Hopkins does not describe the executions as tragic or regrettable. Instead, he treats the convictions and executions as evidence that his methods were effective and that his claim to expertise was justified.

In contrast, the \textit{sadness} identified by the LLM would rather come from Query 5, that does not capture the position of the text. Queries, in this context are rhetorical devices that are then argued against. Given that Hopkins' actively defends his deeds, there is no trace of remorse that would fit the emotion of \textit{sadness}. Instead if anything the driving emotion remains \textit{anger} --- in Hopkins' mind (or at least pen) a justified anger of acting as the sword of god.

Hence, this captures a case where the LLM fails to recover the position of the speaker and the adversarial structure of the source. The example shows how imperfect textual segmentation can interact with failures of speaker attribution. the model identifies salient emotional vocabulary but misinterprets whose attitude that vocabulary represents. While this case does not necessarily  an intrinsic shortcoming of LLMs, much of the mistakes can be attributed to missing context, it constitutes a real challenge in the batch-processing of text as currently employed in the discipline. For further context see \cite{sharpe1996instruments,sharpe2004hopkins} or \cite{healey2024blazing}.

\clearpage
\FloatBarrier

\section{Technical best practise guide}

\subsection{Reporting validation statistics}
\label{app:sec:validation_statistics}

As a first good practise, we always recommend the reporting of validation statistics on a set of test labels after applying a machine learning model. In this section, we outline and explain the standard set of validation statistics commonly used for ML models. However, as argued in section~\ref{subsec:ml_to_replace_human_annotators}, note that even models that perform well on these statistics can still lead to large systematic bias in downstream applications.

The most common set of evaluative statistics are precision rate, recall and F1-statistics. These are based on true positives $TP$, false positives, $FP$, and false negatives, $FN$. They are defined as follows:

\[
\text{Precision} = \frac{TP}{TP + FP},
\]
Precision measures how reliable the set of predicted positives is, i.e.\ the share of predicted positives that are correctly classified. 
\[
\text{Recall} = \frac{TP}{TP + FN},
\]
Recall measures how completely the model recovers the relevant population, i.e.\ the share of true positives that are successfully identified. The F1-statistic combines both metrics as the harmonic mean,
\[
\text{F1} = 2 \cdot \frac{\text{Precision} \cdot \text{Recall}}{\text{Precision} + \text{Recall}},
\]
The F1-staistic is informative when both false positives and false negatives are substantively costly.
These statistics transparently report model performance and are important for judging the appropriateness of a specific ML methodology. In case of unsatisfactory performance researchers are recommended to either change their model or to work on custom built solutions, using e.g. fine-tuning methods (see section~\ref{subsec:scale_of_the_problem}).

\subsection{Practical advice for (near-exact) replicability of LLMs}
\label{app:sec:replicability}

Overall, we recommend the use of locally run models. In the following, we provide a list of best practises to enable as-good-as-possible replications with local models. The current advice was tested using torch and an ollama environment.
\begin{itemize}
    \item When initializing your script
    \begin{itemize}
        \item Fix hardware and threading parameters
        \item Set a seed for the GPU in \textit{torch}
        \item Set a constant number of CPU threads and GPU numbers
        \item Disable parallel evaluation
        \item Disable the \textit{torch} cuDNN auto-tuner that select the fastest algorithm for convolution
    \end{itemize}
    \item In model parameters
    \begin{itemize}
        \item Specify your model not only with the model name but also using sha256 digest - that point to global models that can be pulled by other researchers
        \begin{itemize}
            \item Record the digest in a run manifest
            \item Export your own model and locally save it
        \end{itemize}
        \item Set temperature to zero, set parameters to only select only highest-probability tokens
        \item Set an explicit seed
        \item Fix output length and context parameters
    \end{itemize}
    \item When running the model
    \begin{itemize}
        \item Stabilize GPU kernel selection with warm-up runs
        \begin{itemize}
            \item This forces the GPU backend to complete autotuning and to select consistent kernels
        \end{itemize}
    \end{itemize}
\end{itemize}

Note that these steps should ensure almost perfect replicability. Nonetheless they do not guarantee fully deterministic replicability. Also note that ensuring exact replicability might a complex and demanding task, depending on the concrete hardware and environment the model is run on.

\clearpage

\newpage

\end{document}